\documentclass[journal]{IEEEtran}
\ifCLASSINFOpdf
\else
\fi
\usepackage{bm}
\usepackage{graphicx}
\usepackage{indentfirst}
\usepackage{epsfig}
\usepackage{amsfonts}
\usepackage{amsmath}
\usepackage{url}
\usepackage{color}
\usepackage{algorithm}
\usepackage{algpseudocode}
\usepackage{cancel}
\usepackage{graphicx}
\usepackage{mathcomp}
\usepackage{footnote}
\usepackage{cite}
\usepackage{mathtools}
\newtheorem{theorem}{Theorem}

\newtheorem{proposition}{Proposition}

\usepackage{mathrsfs}
\usepackage{amsfonts}
\usepackage{graphicx}
\usepackage[justification=centering]{caption}
\usepackage{multirow}
\usepackage{hhline}
\usepackage{bigstrut}
\usepackage{amssymb,amsmath}
\usepackage{comment}
\usepackage{nomencl}
\makeglossary

\begin{document}
	\title{Proactive Demand Response for Data Centers: \\A Win-Win Solution}
	
	\author{Hao~Wang,~\IEEEmembership{Student~Member,~IEEE,}
		Jianwei~Huang,~\IEEEmembership{Senior~Member,~IEEE,}\\
		Xiaojun~Lin,~\IEEEmembership{Senior~Member,~IEEE,}
		and~Hamed~Mohsenian-Rad,~\IEEEmembership{Senior~Member,~IEEE}
		
		\thanks{The work is supported by a grant from the Research Grants Council of the Hong Kong Special Administrative Region, China, under Theme-based Research Scheme through Project No. T23-407/13-N. Part of the results have appeared in ACM GreenMetrics 2013 \cite{ourpaper}.}
		\thanks{H. Wang and J. Huang (corresponding author) are with the Network Communications and Economics Lab (NCEL), Department of Information Engineering, The Chinese University of Hong Kong, Shatin, Hong Kong SAR, China, e-mails: \{haowang, jwhuang\}@ie.cuhk.edu.hk.}
		\thanks{X. Lin is with Department of Electrical and Computer Engineering, Purdue University, West Lafayette, IN 47907, USA, e-mail: linx@ecn.purdue.edu.}
		\thanks{H. Mohsenian-Rad is with Department of Electrical Engineering, University of California, Riverside, CA 92521, USA, e-mail: hamed@ee.ucr.edu.}
	}
	\maketitle
	\thispagestyle{empty}  

	\begin{abstract}
		In order to reduce the energy cost of data centers, recent studies suggest distributing computation workload among multiple geographically dispersed data centers, by exploiting the electricity price difference. However, the impact of data center load redistribution on the power grid is not well understood yet. This paper takes the first step towards tackling this important issue, by studying how the power grid can take advantage of the data centers' load distribution \emph{proactively} for the purpose of power load balancing. We model the interactions between power grid and data centers as a two-stage problem, where the utility company chooses proper pricing mechanisms to balance the electric power load in the first stage, and the data centers seek to minimize their total energy cost by responding to the prices in the second stage. We show that the two-stage problem is a bilevel quadratic program, which is NP-hard and cannot be solved using standard convex optimization techniques. We introduce benchmark problems to derive upper and lower bounds for the solution of the two-stage problem. We further propose a branch and bound algorithm to attain the \emph{globally optimal} solution, and propose a heuristic algorithm with low computational complexity to obtain an alternative \emph{close-to-optimal} solution. We also study the impact of background load prediction error using the
		theoretical framework of robust optimization. The simulation results demonstrate that our proposed scheme can not only improve the power grid reliability but also reduce the energy cost of data centers.
	\end{abstract}
	
	\begin{IEEEkeywords}
		Smart grid, data center, demand response, dynamic electricity pricing, load balancing, proactive design.
	\end{IEEEkeywords}
	
	%
	\IEEEpeerreviewmaketitle

\section*{Nomenclature}

\subsection*{Acronyms}
\begin{tabular}{l l} 
	PS1 &Stage-1 problem \\ 
	PS2 &Stage-2 problem \\
	PI &Integrated problem \\
	RS1 &Stage-1 of the restricted problem \\ 
	RS2 &Stage-2 of the restricted problem \\ 
	PE1 &Equivalent problem of the Stage-1 problem \\ 
	PE2 &Equivalent problem of the Stage-2 problem \\ 
	PR1 &Relaxed Stage-1 problem\\
	WCP &Worst-case performance optimization problem
\end{tabular}

\subsection*{Sets}
\begin{tabular}{l l} 
	$\mathcal{T}$ &Set of time slots \\ 
	$\mathcal{N}$ &Set of data centers  
\end{tabular}

\subsection*{Indices}
\begin{tabular}{l l} 
	$t$ &Index of time slots \\ 
	$i$ &Index of data centers
\end{tabular}

\subsection*{Parameters}
\begin{tabular}{l l} 
	 $T$ &Number of time slots \\
	 $N$ &Number of data centers \\
	 $L^{t}$ &Total incoming workload within time slot $t$ \\
	 $M_{i}$ &Total number of servers in data center $i$\\
	 $\mu_{i}$ &Service rate of servers in data center $i$ \\
	 $d_{i}^{t}$ &Transmission delay to data center $i$ in time slot $t$\\ 
	 $D$ &Delay bound\\ 
	 $P_{idle}$ &Average idle power of server \\
	 $P_{peak}$ &Average peak power of server \\
	 $R_{i}$ &Power usage effectiveness of data center $i$\\
	 $\xi$  &Empirical parameter of power consumption\\
	 $\alpha_{i}^{t}$ &Base price for data center $i$ in time slot $t$\\
	 $\beta_{i}$ &Sensitivity parameter of price for data center $i$\\ 
	 $Q_{i}^{t}$ &Available supply to data center $i$ in time slot $t$\\
	 $B_{i}^{t}$ &Background load in location $i$ and time slot $t$\\ 
	 $C_{i}$ &Power capacity in location $i$ \\
	 $\underline{\pi}_{i}^{t}$ &Price lower bound for data center $i$ in time slot $t$\\
	 $\overline{\pi}_{i}^{t}$ &Price upper bound for data center $i$ in time slot $t$\\
	 $\pi_{\max}^{t}$ &Maximum average price in time slot $t$ \\
	 $\theta_{i}$ &Coefficient for energy consumption of data center $i$ \\
	 $E^{t}$ &Total energy required in time slot $t$ \\
	 $\underline{E}_{i}^{t}$ &Energy lower bound for data center $i$ in time slot $t$ \\
	 $\overline{E}_{i}^{t}$ &Energy upper bound for data center $i$ in time slot $t$ \\
	 $\Delta_{i,\min}^{t}$ &Error lower bound in location $i$ and time slot $t$ \\
	 $\Delta_{i,\max}^{t}$ &Error upper bound in location $i$ and time slot $t$
\end{tabular}

\subsection*{Variables}
\begin{tabular}{l l} 
	$\lambda_{i}^{t}$ &Workload assigned to data center $i$ in time slot $t$\\
	$x_{i}^{t}$ &Number of active servers in data center $i$ and time $t$\\ 
	$e_{i}^{t}$ &Energy consumption of data center $i$ in time slot $t$\\ 
	$s_{i}^{t}$ &Billing reference for data center $i$ in time slot $t$\\ 
	$\pi_{i}^{t}$ &Unit energy price for data center $i$ in time slot $t$\\
	$r_{i}^{t}$ &Electric load ratio in location $i$ and time slot $t$\\
	$\underline{z}_{i}^{t}$ &Binary variables \\
	$\overline{z}_{i}^{t}$ &Binary variables \\
	$\delta_{i}^{t}$ &Load prediction error in location $i$ and time slot $t$
\end{tabular}

	\section{Introduction}
	Energy management of large and distributed data centers has become an increasingly important problem. With the fast development of cloud computing services, it is now common for a cloud provider (\textit{e.g.}, Google, Microsoft, and Amazon) to build multiple, large, and geographically dispersed data centers across the continent. Each data center may include hundreds of thousands of servers, massive storage equipment, cooling facilities, and power transformers. The energy consumption and cost of data centers hence can be significant \cite{sigcomm}. For example, Google reported in 2011 that its data centers continuously draw almost 260 MW of power, which is more than what Salt Lake City consumes \cite{google}. Microsoft's data center in Washington US consumes 48 MW of power, which is equivalent to the power consumption of about 40,000 households. This has motivated growing research activities toward optimizing the data center operations to reduce the total energy cost. For example, Qureshi \emph{et al.} in \cite{sigcomm} proposed an energy cost minimization method for distributed data centers to exploit electricity price difference. The idea is later extended in \cite{infocom,sigmetrics,online,SLA,price,Auction,tradeoff}.
	
	However, most existing studies of energy management of distributed data centers have focused on the energy cost minimization from the viewpoint of data centers, but fail to consider the impact of such energy management practice on the power grid. Note that, due to their enormous energy consumption, data centers are expected to have a great influence on the operation of the power grid \cite{coordinate}. Without taking such impact into account, these energy management schemes may adversely affect power-grid stability and load balancing.
	
	In this paper, we aim to study the energy cost minimization of distributed data centers based on their impact to the power grid. We seek to benefit from the recent advances in two-way communications that are available in smart grid \cite{smartgrid} to allow interactions and coordinations between energy suppliers and consumers in real time to improve demand side management. In our proposed framework, the utility company can set dynamic prices to the \emph{demand-responsive data centers}, and the data centers can dynamically change energy consumption in response to the price changes. This can effectively coordinate demand with supply, and hence avoid unintended power overloading.
	
\begin{figure}[t]
		\centering
		\includegraphics[width=8.5cm]{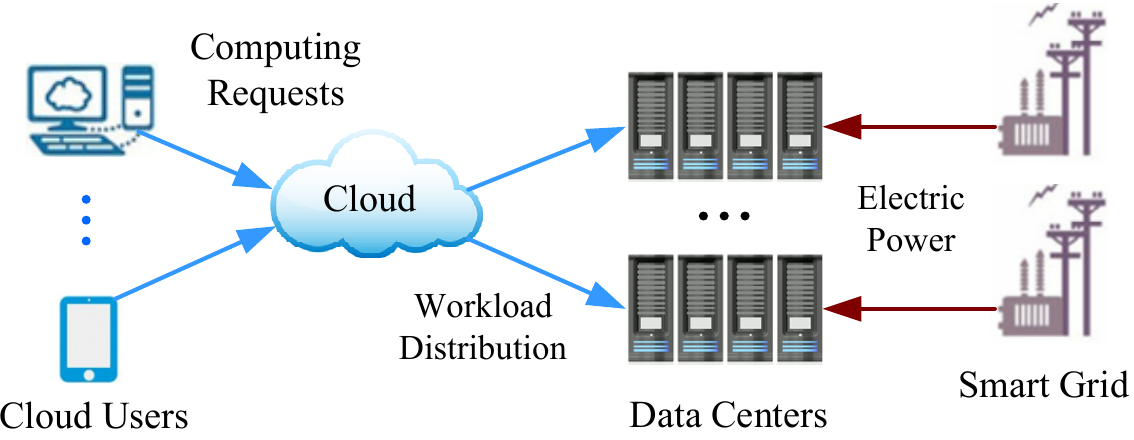}
		\caption{\label{fig_interaction}Smart grid and data center interaction.}
	\end{figure}

	The overall framework of our proposed system setup is shown in Fig. \ref{fig_interaction}. Cloud service users send computing requests via Internet to the cloud provider. Exploiting various electricity prices at different locations, the cloud provider minimizes the total energy cost by assigning users' requests to different data centers. The utility company utilizes the demand response of data centers, and tries to achieve power load balancing by altering the electricity consumption of data centers through dynamic pricing.
		
	The main contributions of this paper are as follows:
	\begin{itemize}
		\item \textit{Data center and smart grid interaction}: To the best of our knowledge, this is the first paper that studies the interactions between smart grid and data centers by considering the active decisions on both sides. In particular, how does the utility company properly incentivize data centers to provide demand response services toward a reliable power grid?
		\item \textit{Modeling and solution methods}: We formulate the interactions between smart grid and data centers as a two-stage price optimization problem. In its original form, this problem cannot be solved by standard convex programming techniques. Therefore, we reformulate the problem as a mixed integer quadratic program, and design a customized branch-and-bound algorithm to attain the globally optimal solution. We also design a low-complexity descent algorithm to attain a close-to-optimal solution.
		\item \textit{Performance benchmarks}: To help characterizing the optimal solution of the two-stage price optimization problem, we construct two single-level optimization problems, namely an \emph{Integrated Problem} and a \emph{Restricted Problem}, which correspond to the performance upper and lower bounds of the two-stage price optimization problem.
		\item \textit{Case studies and implications}: Our proposed method can not only balance the power load for smart grid but also reduce total energy cost for data centers, hence achieving a \emph{win-win} result. 
	\end{itemize}

	The remainder of the this paper is organized as follows. We review the related work in Section II. After that, we formulate the system model as a two-stage price optimization problem in Section III. In Section IV, we study two benchmark problems to provide performance bounds for the formulated two-stage price optimization problem. In Section V, we analyze the solution of the two-stage price optimization problem, design a branch-and-bound algorithm to yield the global optimum, and propose an alternative heuristic algorithm to solve the sub-optimal solution. In Section VI, we analyze the worst-case performance by considering the prediction error in background power load. Performance of the proposed scheme is evaluated in Section VII. This paper is concluded in Section VIII.
	
	\section{Related Work and Motivation}
	\subsection{Literature Review}
	There are many existing research results on managing data center's workload to reduce energy cost, such as those studying the energy cost minimization problem with multi-electricity-market environment \cite{infocom}, green renewable generators \cite{sigmetrics}, online optimization \cite{online}, service level agreements \cite{SLA}, and deregulated electricity price \cite{price}. Zhang \textit{et al.} \cite{Auction} designed a Vickrey-Clarke-Groves auction mechanism, in which tenants of data centers voluntarily bid for emergency demand response. However, these results did not consider the active response by the utility companies, nor did they consider how the data centers' demand response may bring large load fluctuations across different locations over time. This motivates us to study the interactions between smart grid and geographically dispersed data centers, and examine how smart grid can properly incentivize data centers through dynamic pricing to improve the grid reliability.
	
	There has been a large body of research on demand response of strategic energy consumers \cite{hamed1,han,lina,mung}. For example, in \cite{hamed1}, Mohsenian-Rad and Leon-Garcia suggested scheduling household devices based on the predicted prices to minimize the electricity cost. In \cite{han}, Nguyen \textit{et al.} proposed a game theoretic model, in which an electricity provider dynamically updates the energy prices to reduce the peak load, by considering the load profiles of users. In \cite{lina}, Li \textit{et al.} studied demand response based on utility maximization, and proposed a distributed algorithm to compute optimal prices and power schedules. In \cite{mung}, Wong \textit{et al.} designed a time-dependent price to incentivize users to shift power load so as to relieve stress during peak hours. 
	
	\subsection{Motivation}
	Different from traditional residential or industrial consumers, data centers are special electricity consumers. This is not only because of their enormous energy consumption, but also because of flexibility of energy consumptions over multiple locations. The previous studies in \cite{infocom,sigmetrics,online,SLA,price,Auction} mainly focused on the workload distribution from the perspective of data centers. As reported in \cite{tradeoff}, such workload distribution of data centers has great impact on power load balancing in the smart grid. 
		
	In the power system, the utility company is responsible for supplying power to meet the demand, and for maintaining the safe operation of the smart grid system. The utility company can utilize the demand response of data centers to manage their energy consumption. However, most of the existing demand response programs focused on the time flexibility of residential demands, without considering the demand side management over multiple locations. The latter is difficult to do for residential demands, but is very suitable in the case of geographically dispersed data centers.\footnote{The cloud provider owns multiple data centers located in different geographical locations, and thus gains flexibility of power loads over locations via workload assignment over different data centers. As an example, when Google responds to a user's web search query, the corresponding computation can be done in any of the Google's data centers (as long as certain service quality agreement is satisfied).} This motivates us to design the dynamic pricing incentive mechanism from the grid operator's point of view, in order to incentivize the proper demand response from multiple geographically dispersed data centers. Tran \textit{et al.} \cite{MultiDC} studied demand response of data centers in a multi-utilities environment, and modeled the interactions between utilities and data centers as a Stackelberg game. Different from \cite{MultiDC}, we study the interaction between one utility company and one cloud provider (with multiple data centers) as a bi-level optimization problem, propose two benchmark problems to estimate the performance bounds, and propose two algorithms to solve the optimal prices and close-to-optimal prices, respectively.

	\section{System Model}
	We consider a discrete time model $t \in \mathcal{T} = \{ 1,...,T \}$, where the length of a time slot matches the time-scale at which the workload allocation decisions and dynamic pricing decisions are updated, \emph{e.g.} once an hour \cite{infocom}. Let $\mathcal{N}=\{1,...,N\}$ denote the set of geographically dispersed data centers, where each data center $i \in \mathcal{N}$ has $M_{i}$ homogeneous servers, and has the same function in terms of supporting various kinds of applications (e.g., Internet services, image processing). As we will explain later, not all servers are turned on during each time slot. 
	
\begin{figure}[t]
		\centering
		\includegraphics[width=8.5cm]{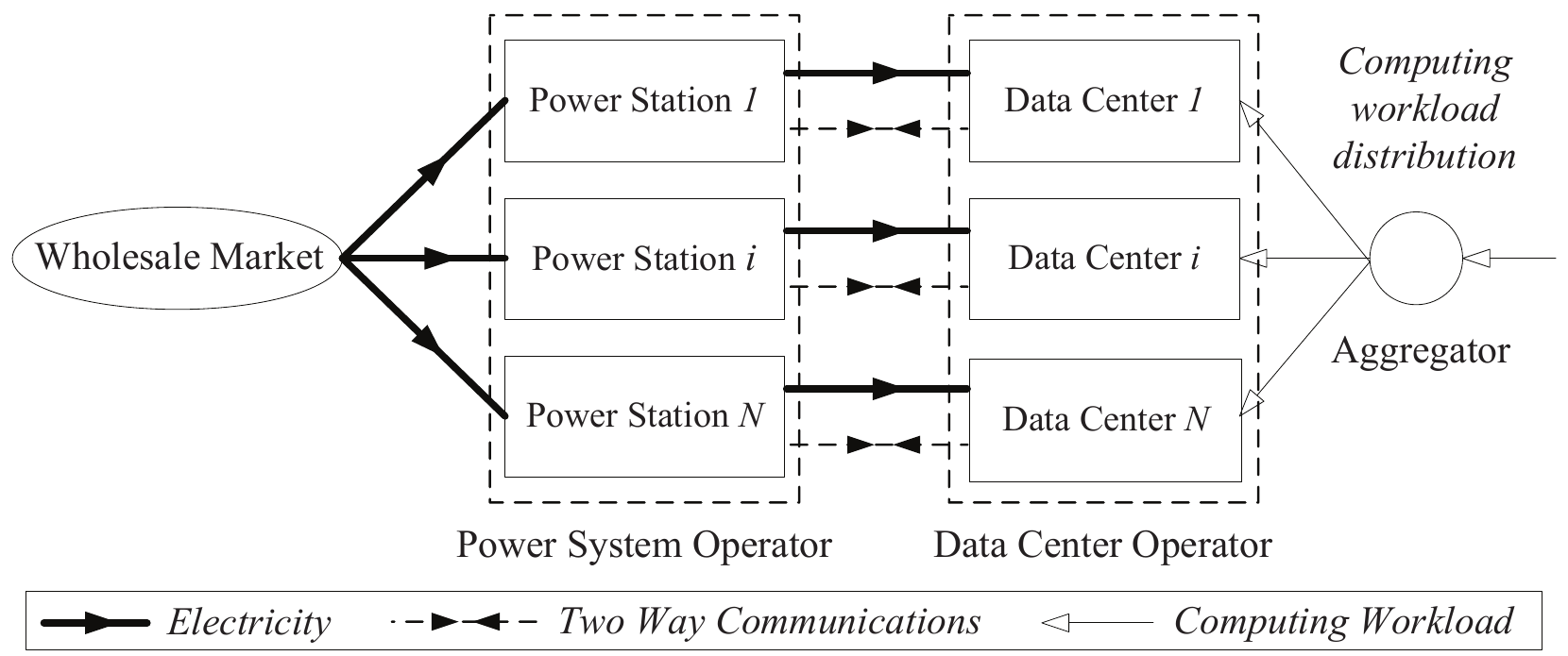}
		\caption{\label{fig_architecture}The architecture for data center demand response.}
	\end{figure}

	Fig. \ref{fig_architecture} illustrates the system architecture of data centers and smart grid. We assume that a group of geographically dispersed data centers are operated by a single cloud provider, and there is a traffic aggregator (\emph{e.g.}, a front-end portal server) responsible for distributing the total incoming computing workload $L^{t}$ within time slot $t$ to data centers in different regions \cite{infocom}. Each data center is powered by a dedicated power substation in the power grid, and all the substations are operated by the same utility company.\footnote{Many practical examples motivates our assumption of one utility company. For example, Alibaba cloud, a Chinese cloud provider, runs five data centers at different locations in China, and three of which are served by the State Grid Corporation of China. Such scenarios also exist in deregulated electricity markets, such as in California US \cite{California}.}
	In each time slot $t$, we model the interactions between utility company and data centers in two stages. In Stage 1, the utility company sets a billing reference $s_{i}^{t}$, which determines the electricity tariff as we will explain later for each data center $i$ to balance the load on the power grid. In Stage 2, we assume that the data centers can predict the workload accurately at the beginning of each time slot. Then data centers cooperate with each other (as they belong to the same cloud operator) so as to minimize the total energy cost by determining the computing workload allocation $\lambda_{i}^{t}$ and the number of active servers $x_{i}^{t}$ in each data center $i$. Next, we discuss these decisions in details.

	\subsection{Stage 2: Data Center's Energy Cost Minimization}
	First, we consider the Stage-2 problem, where a cloud provider (such as Google) wants to minimize the total energy cost of multiple data centers. In practice, data centers directly negotiate with the utility company regarding the electricity rates \cite{contract}. In time slot $t$, the utility company charges data center $i$ with the following regional electricity price $\pi_{i}^{t}$ per unit of energy:
	\begin{align}\label{unitprice}
		& \pi_{i}^{t}=\alpha_{i}^{t} +\beta_{i} (e_{i}^{t}-s_{i}^{t}),
	\end{align}
	where $e_{i}^{t}$ is the data center's the electricity consumption, $s_{i}^{t}$ is called the billing reference, $\beta_i>0$ is a sensitivity parameter, and $\alpha_{i}^{t}>0$ denotes the base price, all at location $i$ in time slot $t$. The dynamic pricing scheme in \eqref{unitprice} is motivated by the tiered electricity pricing, which has been widely implemented in various power markets such as the United States, Japan, and China. The key idea of tiered pricing is to set several pricing tiers for the energy consumption, and the unit price per unit of energy increases with the tiers progressively \cite{tier}. In \eqref{unitprice}, the term $\beta_{i} (e_{i}^{t}-s_{i}^{t})$ reflects the difference between electricity consumption $e_{i}^{t}$ and the billing reference $s_{i}^{t}$. The unit price $\pi_i^t$ will be higher than the base price if $e_{i}^{t} > s_{i}^{t}$.
	
	Next, we discuss the data centers' optimization constraints.
	
	\subsubsection{Workload constraint}
	In each time slot $t$, users' computing requests (workload to the cloud provider) are received by a front-end portal server. Then a total of $N$ data centers should work together to complete the total workload of $L^t$, with the allocation to data center $i$ as $\lambda_i^t$:
	\begin{align}\label{workload}
		\sum_{i=1}^{N} \lambda_{i}^{t}=L^{t},~\lambda_{i}^{t} \geq 0, ~\forall i \in \mathcal{N}, t \in \mathcal{T}.
	\end{align}
	
	\subsubsection{QoS (delay) constraint}
	It is important for data centers to provide QoS guarantees to the users, and QoS can be specified by the service level agreement (SLA) \cite{sla}. SLA usually measures the average performance for the operation of a data center during a period of time. We consider both the \emph{transmission delay} (incurred before the request arrives at a data center) and the \emph{queuing delay} (experienced after the request arrives at a data center). We define $d_{i}^{t}$ as the transmission delay experienced by a computing request from the aggregator to data center $i$ during time slot $t$. Notice that $d_i^t$ is usually much less than the length of a time slot. To model the queuing delay, we use queuing theory to estimate the average processing time in data center $i$ when there are $x_{i}^{t}$ active servers processing workload $\lambda_{i}^{t}$ with a service rate $\mu_{i}$ per server.\footnote{We assume that the servers in the same data center $i$ are homogeneous and have the same service rate $\mu_i$.} Applying the results from M/M/1 queuing theory \cite{sigmetrics}, the average waiting time is approximately $\frac{1}{\mu_i x_{i}^{t}- \lambda_{i}^{t}}$. To meet the QoS requirement, the total time delay experienced by a computing request should satisfy some delay bound $D$, which is the maximum waiting time that a request can tolerate. For simplicity, in this paper, we will assume homogeneous requests that have the same delay bound $D$. Therefore, we have the following QoS constraint
	\begin{align}\label{qos}
		d_{i}^{t} + \frac{1}{\mu_i x_i^t- \lambda_i^t} \leq D,~\forall i \in \mathcal{N}, t \in \mathcal{T},
	\end{align}
	where $\mu_i x_{i}^{t} > \lambda_i^t$.
	
	\subsubsection{Server constraint}
	At each data center $i$, there are tens of thousands of servers providing cloud computing services to meet users' requests. Let $M_{i}$ denote the maximum number of available servers. The cloud provider can switch on and off servers to adjust the service time. Since the number of servers is usually large, we can relax the integer constraint on the number of active servers without significantly affecting the optimal result. Therefore, we have the following server constraint\footnote{We set the minimum required number of active servers in each data center as zero. It can also be set as a non positive to reflect operational requirements for the data center, without changing the engineering insights from the analysis.}
	\begin{align}\label{server}
		0 \leq x_{i}^{t} \leq M_{i},~\forall i \in \mathcal{N}, t \in \mathcal{T}.
	\end{align}
	
	\subsubsection{Energy consumption constraint}
	The energy consumption of data centers consists of IT energy consumption (\emph{e.g.}, CPU, memory, and storage) and ancillary energy consumption (\emph{e.g.}, cooling, lighting, and power facility). The quantitative relation between IT energy consumption and ancillary energy consumption is measured by the \textit{power usage efficiency} (PUE) \cite{pue}, which is defined as the ratio of total energy consumption to IT energy consumption. The energy used by computing equipments is considered to be productive. On the contrary, the energy for ancillary infrastructure (\emph{e.g.}, cooling, lighting, and power facility) is auxiliary. PUE helps us understand the total energy consumption based on the IT energy consumption. Therefore, we can calculate the total energy consumption of a data center using PUE, amount of computing workload, and number of active servers. Precisely, based on the data center power model in \cite{coordinate}, we formulate the energy consumption of data center $i$ in time slot $t$ as
	\begin{align*}
		& e_{i}^{t} = x_{i}^{t} \left( P_{idle} + (R_{i} -1)  P_{peak} \right)+ x_{i}^{t} (P_{peak}-P_{idle}) \gamma_{i}^{t} + \xi_i,
	\end{align*}
	where $P_{idle}$ and $P_{peak}$ represent the average idle power and average peak power of a single server, respectively. The power efficiency parameter $R_{i} > 1$ denotes PUE of data center $i$. The parameter $\xi_i$ is an empirical constant indicating the base energy consumption of data center $i$, and $\gamma_{i}^{t}$ denotes the average server utilization of data center $i$ in time slot $t$. 
	
	We substitute the average server utilization $\gamma_{i}^{t}=\lambda_{i}^{t} / (\mu_i x_{i}^{t})$, and rewrite $e_i^t$ in the following equivalent form:
	\begin{equation}\label{energycon}
	  \begin{split}
     e_{i}^{t} =  \left( P_{idle} + (R_{i} -1)  P_{peak} \right) x_{i}^{t}   
      + \frac{P_{peak}-P_{idle}}{\mu_i} \lambda_{i}^{t} + \xi_i , \\
    \forall i \in \mathcal{N}, t \in \mathcal{T}, 
	  \end{split}
	\end{equation}
	which is an affine function with respect to the number of active servers $x_{i}^{t}$ and the computing workload $\lambda_{i}^{t}$.
		
	Given the operational requirements of the power substation, we limit the maximum power that can be consumed by data center $i$ in time slot $t$ as
	\begin{equation}\label{maxenergy}
	0 \leq e_{i}^{t} \leq Q_{i}^{t},~\forall i \in \mathcal{N}, t \in \mathcal{T}.
	\end{equation}
	where $Q_{i}^{t}$ denotes the available power supply to data center $i$ in time slot $t$.
	
	With the above constraints, we can formulate the cloud provider's energy cost minimization problem in Stage 2. The objective is to minimize the data centers' total energy cost over all locations and all time slots by choosing the workload allocation $\lambda_{i}^{t}$ and the number of active servers $x_{i}^{t}$ for each data center $i\in\mathcal{N}$ and each time $t\in\mathcal{T}$. As the operational constraints \eqref{workload}-\eqref{maxenergy} are decoupled across time slots, we formulate the energy cost minimization problem in time slot $t$ as follows:
	\begin{align*}
	& \leftline{\textbf{Stage-2 Problem (PS2): Total Energy Cost Minimization}}
	\end{align*}
	\begin{equation*}
	  \begin{aligned}
	  & \min_{\boldsymbol{\lambda}^{t},~\boldsymbol{x}^{t}} 
	  && \sum_{i \in \mathcal{N}} \Big( \alpha_{i}^{t} +\beta_{i} (e_{i}^{t}-s_{i}^{t}) \Big) e_{i}^{t} \\
	  & \text{subject to}  
	  && \text{Constraints \eqref{workload}--\eqref{maxenergy}},
	  \end{aligned}
	\end{equation*}
	where $\boldsymbol{\lambda}^{t} = \{ \lambda_i^t,~\forall i\in\mathcal{N}\}$ and $\boldsymbol{x}^{t} = \{ x_i^t,~\forall i\in\mathcal{N} \}$ denote the workload allocation vector and active server number vector for each time slot $t\in\mathcal{T}$, respectively. The energy cost of data center $i$ is calculated as the product of its energy consumption $e_{i}^{t}$ and the corresponding unit price $\alpha_{i}^{t} +\beta_{i} (e_{i}^{t}-s_{i}^{t})$.
	
	Note that, the optimal value of workload allocation $\lambda_i^t$, number of active servers $x_i^t$ and energy consumption $e_i^t$ in \eqref{energycon} are functions of the billing references $\boldsymbol{s}^{t}=\{ s_i^t,~\forall i\in\mathcal{N} \}$ in time slot $t$. Given $\boldsymbol{s}^{t}$, we can solve Problem \textbf{PS2}, and will present the optimal solutions of $\lambda_i^t$, $x_i^t$ and $e_i^t$ in Section V.

	\subsection{Stage 1: Smart Grid's Power Load Balancing Problem}
	We are now ready to consider the Stage-1 power load balancing problem for the smart grid. We classify the load into two groups: data centers and others. We focus on the data centers' loads as they have geographical flexibility, and let the latter group as background loads. With the emergence of smart grid communications technologies, it is possible for the utility company to incentivize the data centers to shift loads from heavily loaded regions to lightly loaded regions. In our proposed framework, the smart grid optimizes dynamic tiered prices by setting the billing references $\boldsymbol{s}^{t}$ in each time slot $t$ to balance power load across geographical locations. To measure the power load levels in different locations, we define the \emph{electric load ratio} in location $i$ and time slot $t$ as
	\begin{align}\label{loadratio}
		r_{i}^{t}(\boldsymbol{s}^{t}) = \frac{e_{i}^{t}(\boldsymbol{s}^{t}) + B_{i}^{t}} {C_{i}},
	\end{align}
	where $B_{i}^{t}$ is the background power load, and $C_{i}$ is the capacity of power substation $i$. Note that the load ratio $r_{i}^{t}$ is a function of the energy consumption $e_{i}^{t}$, and thus also depends on the billing reference $\boldsymbol{s}^{t}$ for all locations in time slot $t$. The utility company aims at balancing the load ratio $r_{i}^{t}(\boldsymbol{s}^{t})$ at all locations in each time slot.
	
	Let $Q_{i}^{t} = C_{i} - B_{i}^{t}$ be the maximum available power supply to data center $i$ in time slot $t$. Since our study focuses on the demand response of data centers, we denote the aggregate energy usage of all the users other than data centers as the background energy load. We assume that the utility company is able to accurately forecast\footnote{We first solve the two-stage problem assuming perfect background load prediction. In section VI, we will further study the impact of prediction error.} the background energy load ahead of each time slot \cite{predict}.
	
	Based on the load ratio $r_{i}^{t}$, we define the \emph{electric load index (ELI)} across all locations in time slot $t$ as
	\begin{align}\label{def_eli}
		ELI \triangleq \sum_{i \in \mathcal{N}}
		~ \Big( r_{i}^{t}(\boldsymbol{s}^{t}) \Big) ^{2} C_{i},
	\end{align}
	where \emph{ELI} measures the overall load ratio across all locations. Note that electric load ratio $r_{i}^{t}$ is a normalized indicator, which does not reflect the importance of those locations with large capacities. Therefore, we introduce the capacities $C_{i}$ as the weighted coefficients in \emph{ELI}. We can show that minimizing \emph{ELI} with respect to $e_{i}^{t}$  yields an equal load ratio across all locations in the ideal case (without considering any constraints):
	\begin{align*}
	\frac{e_{1}^{t}+B_{1}^{t}}{C_{1}} = \cdots = \frac{e_{N}^{t}+B_{N}^{t}}{C_{N}},
	\end{align*}
	which indicates no overloading problem occurs in any of the locations. Therefore, the system reliability is improved at these locations.
	
	However, such even load distribution may not be achievable in practice, because the energy consumption $e_{i}^{t}$ should also satisfy the operational constraints for workload allocation and number of active servers in \eqref{workload}--\eqref{maxenergy}. Moreover, the cloud provider and the utility company are independent entities. Data centers are operated by the cloud provider, which implies that the energy consumption of data centers cannot be directly controlled by the utility company.

	In order to balance the electricity load, in this paper we focus on the scenario where the utility company charges dynamic prices to incentivize users to shift their electricity usage to less loaded locations. To encourage the participation of data centers into the demand response program and prevent the utility company from abusing its market power, constraints should be set to regulate the dynamic prices. In practice, the utility company and data centers usually negotiate with each other and enter into a contract \cite{contract} to specify the pricing structure. Based on related studies \cite{energyecon}, we set the following constraints for the energy price $\pi_i^t$:
	\begin{align}
		& \underline{\pi}_i^t \leq \alpha_{i}^{t} + \beta_{i} (e_{i}^{t}-s_i^t) \leq \overline{\pi}_i^t,~\forall i \in \mathcal{N}, t \in \mathcal{T},  \label{s_constraint1} \\
		& \frac{1}{N} \sum_{i \in \mathcal{N}} \left[ \alpha_{i}^{t} + \beta_{i} (e_{i}^{t}-s_i^t) \right] \leq \pi_{\max}^{t},~t \in \mathcal{T}, \label{s_constraint2}
	\end{align}
	where \eqref{s_constraint1} ensures that the price charged to the data centers is always contained within the range $[\underline{\pi}_i^t,\overline{\pi}_i^t]$. Constraint \eqref{s_constraint2} enforces that the dynamic prices across all locations have an average price ceiling $\pi_{\max}^{t}$, which is specified by the contract between the utility company and data centers \cite{energyecon}. Constraint \eqref{s_constraint2} can prevent the utility company from charging the maximum possible price in all locations. More precisely, the utility company has to provide lower prices to other locations if it charges a higher prices at some locations, so that \eqref{s_constraint2} can be satisfied. This will give a guarantee to the cloud provider, such that the dynamic price will not arbitrarily increase the energy cost of the data centers.
	
	After the contract terms (\emph{e.g.}, constraints \eqref{s_constraint1} and \eqref{s_constraint2}) are settled, the utility company is responsible of enforcing the price constraints \eqref{s_constraint1} and \eqref{s_constraint2}.\footnote{To enforce constraints \eqref{s_constraint1} and \eqref{s_constraint2}, the utility company should carefully determine the dynamic prices and consider the corresponding responses from the data centers, as the price constraints \eqref{s_constraint1} and \eqref{s_constraint2} involve both dynamic prices and energy consumption responses of data centers.} We formulate the smart grid's load balancing problem in time slot $t$ as follows:
	\begin{align*}
		\leftline{\textbf{Stage-1 Problem (PS1): Electric Power Load Balancing}}
	\end{align*}
	\begin{equation*}
		\begin{aligned}
			& \underset{\boldsymbol{s}^{t}} {\min}
			& & \sum_{i \in \mathcal{N}}~~ \Big( r_{i}^{t}(\boldsymbol{s}^{t}) \Big) ^{2} C_{i} \\
			& \text{subject to}
			& & \text{Constraints \eqref{s_constraint1} and \eqref{s_constraint2}},
		\end{aligned}
	\end{equation*}
	where the electric load ratio $r_{i}^{t}$ depends on the energy consumption $e_{i}^{t}$, which is the optimal solution of Stage-2 Problem \textbf{PS2}.
	
	\subsection{Two-stage Price Optimization Problem}
	For Problem \textbf{PS2}, we can show that constraints \eqref{workload}--\eqref{maxenergy} can be equivalently rewritten as constraints of data centers' energy consumption:
	\begin{align}
	& \sum_{i \in \mathcal{N}} \theta_i e_{i}^{t} = E^{t}, \label{energy_constraint1} \\
	& \underline{E}_{i}^{t} \leq e_{i}^{t} \leq \overline{E}_{i}^{t},~\forall i\in\mathcal{N}  \label{energy_constraint2},
	\end{align}
	where $\theta_i$, $E^{t}$, $\underline{E}_{i}^{t}$ and $\overline{E}_{i}^{t}$ are system parameters. Constraint \eqref{energy_constraint1} is derived from the workload constraint in \eqref{workload}, which specifies that the summation of $\theta_i$-weighed energy consumption of all the data centers should reach $E^{t}$ in order to process the total workload $L^{t}$. The box constraint \eqref{energy_constraint2} sets the energy consumption upper bound $\underline{E}_{i}^{t}$ and lower bound $\overline{E}_{i}^{t}$ for each data center, to meet all the inequality constrains in \eqref{qos}--\eqref{maxenergy}. For the proof and detailed representation of the parameters, please see Appendix A.
	
	Using constraints \eqref{energy_constraint1} and \eqref{energy_constraint2}, we can simplify Problem \textbf{PS2} into an equivalent energy consumption distribution problem, in which the cloud provider directly decides the energy consumption of data center $e_i^t$ to minimize the energy cost. The equivalent energy consumption distribution problem is presented as follows:
    \begin{align*}
	\leftline{\textbf{PE2: Equivalent Problem of PS2}}
	\end{align*}
	\begin{equation*}
	\begin{aligned}
	& \underset{\boldsymbol{e}^{t}} {\min}
	& & \sum_{i \in \mathcal{N}} 
	\Big( \alpha_{i}^{t} +\beta_{i} (e_{i}^{t}-s_{i}^{t}) \Big) e_{i}^{t} \\
	& \text{subject to}
	& & \text{Constraints \eqref{energy_constraint1} and \eqref{energy_constraint2}},
	\end{aligned}
	\end{equation*}
	where $\boldsymbol{e}^{t}= \{ e_i^t,~\forall i \in \mathcal{N} \}$. Once the energy consumption $e_i^t$ is determined, we can find the corresponding workload allocation $\lambda_i^t$ and number of active servers $x_i^t$.
		
	\begin{figure}[t]
		\centering
		\includegraphics[width=9cm]{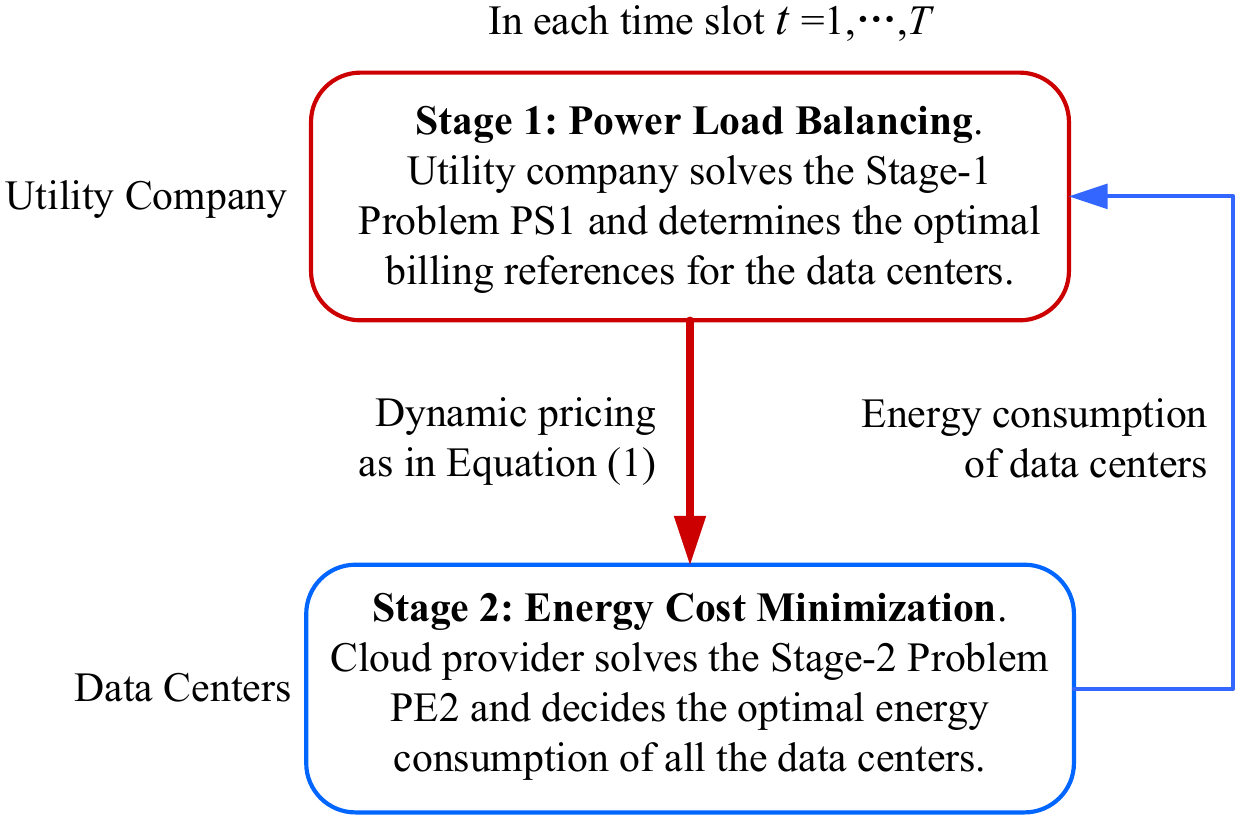}
		\caption{\label{fig_twostage}Two-stage optimization problem.}
	\end{figure}	
	Fig. \ref{fig_twostage} shows the relation between the two-stage problems \textbf{PS1} and \textbf{PE2}, each of which is executed once in each time slot. In Stage 1, at the beginning of each time slot, the utility company sets billing references for data centers to optimize the \emph{ELI} performance. This leads to the tiered price $\pi_i^t = \alpha_i^t + \beta_i (e_i^t -s_i^t)$ for each data center $i$. In Stage 2, the cloud provider optimizes the energy consumption $e_i^t$ of each data center in order to minimize the total energy consumption $\sum_{i\in\mathcal{N}} \pi_i^t e_i^t$ in time slot $t$.
		
	The two-stage problem is a challenging optimization problem to solve, due to the coupled variables and constraints. As the utility company aims to balance the electric load across locations, it will consider the response of the cloud provider in Stage 2, when computing the optimal billing references $\boldsymbol{s}^t$ in Stage 1. Before solving the two-stage problem, we will introduce two benchmark problems to bound the optimal solution.

	\section{Performance Benchmarks}
	The two-stage problem is a quadratic bilevel program with coupled constraints, which is NP-hard in general and cannot be solved effectively by standard convex optimization algorithms. Before proposing solution methods to solve the two-stage problem, we construct two benchmarks, the integrated problem and the restricted problem, to provide lower bound and upper bound of the \emph{ELI} performance, which are helpful in terms of solving the two-stage problem in Section V.

	\subsection{The Integrated Problem}
	We consider the following integrated problem as a benchmark, where the utility company directly decides the optimal workload assignments and the number of active servers for each data center (without the need of dynamic pricing). This will reveal the minimum \emph{ELI} that the system can achieve if the utility company and the data centers fully cooperate with each other.
	
	The integrated problem is formulated as follows.
	\begin{align*}
		\leftline{\textbf{PI: Integrated Problem}}
	\end{align*}
	\begin{equation*}
		\begin{aligned}
			& \underset{\boldsymbol{\lambda}^{t},~\boldsymbol{x}^{t}} {\min}
			& & \sum_{i \in \mathcal{N}}
			~~ \Big( r_{i}^{t} \Big) ^{2} C_{i} \\
			& \text{subject to}
			& & \text{Constraints \eqref{workload}--\eqref{maxenergy}}.
		\end{aligned}
	\end{equation*}
	
	We can see that the objective is consistent with the utility company's objective of load balancing in Problem \textbf{PS1}. The constraints are the same in Problem \textbf{PS2} for data centers' operation. Problem \textbf{PI} is a convex quadratic program, which can be solved by standard convex optimization techniques \cite{convex}. 
	
	Intuitively, compared with the scenario where the utility company incentivizes data centers through dynamic pricing, direct control of data centers' operation would be more efficient in terms of load balancing. This can lead to a lower bound of the \emph{ELI} performance stated in the following proposition.
	\begin{proposition}\label{lowerbound}
		The optimal solution of the integrated problem \textbf{PI} provides a lower bound of the optimal \emph{ELI} performance of \textbf{PS1}.
	\end{proposition}
	
	To prove \emph{Proposition 1}, we need to the show that the feasible set of the integrated problem \textbf{PI} is larger than that of the original two-stage optimization problem. In the integrated problem \textbf{PI}, the utility company directly controls the workload allocation and the number of servers in the data centers, subject to the data center operation constraints \eqref{workload}-\eqref{maxenergy}. Whereas in the two-stage problems \textbf{PS1} and \textbf{PS2}, the utility company aims at indirectly managing data centers' operation in \textbf{PS2} through price incentives in \textbf{PS1}, subject to both data center operation constraints \eqref{workload}-\eqref{maxenergy} and pricing constraints \eqref{s_constraint1}-\eqref{s_constraint2}. Intuitively, when the utility company directly controls data centers' operation in \textbf{PI}, the decision is more flexible than incentive-based in through the two-stage problems \textbf{PS1} and \textbf{PS2}. Hence, the performance of \textbf{PI} should be better, which means a lower \emph{ELI}. For detailed proof, see Appendix B.
	
	Note that the \emph{ELI} performance gap between the estimated lower bound and the optimal solution is affected by constraints \eqref{s_constraint1} and \eqref{s_constraint2} in the two-stage problem. For example, enlarging the price range $[\underline{\pi}_i^t,\overline{\pi}_i^t]$ in price constraint \eqref{s_constraint1} can improve the optimal \emph{ELI} performance to be close to the \emph{ELI} lower bound, as the dynamic pricing scheme of the utility company has a larger feasible set.

	\subsection{The Restricted Problem}
	After we provide a lower bound for \emph{ELI} by solving \textbf{PI}, we present another benchmark problem namely restricted problem \textbf{RS}. 
	
	In order to construct the restricted problem, first, we note that in the two-stage problem, different stages have different constraints that cannot be moved across stages. The Stage-1 problem is the upper-level problem, while the Stage-2 problem is the lower-level problem. The constraints of the Stage-1 problem \textbf{PS1} also apply to the Stage-2 problem \textbf{PE2}, but the operational constraints of data centers in the Stage-2 problem \textbf{PE2} only need to be satisfied by the data centers. Intuitively, moving constraints from the Stage-2 problem to the Stage-1 problem shrinks the utility company's action set. Thus, the way we formulate the restricted problem is to move the bounding constraint on energy consumption \eqref{energy_constraint2} in \textbf{PE2} to the Stage-1 problem \textbf{PS1}. Thus we formulate the restricted problem in time slot $t$ with the modified Stage-1 and Stage-2 problems as follows. 
	\begin{align*}
		\leftline{\textbf{RS1: Stage 1 of the Restricted Problem}}
	\end{align*}
	\begin{equation*}
		\begin{aligned}
			& \underset{\boldsymbol{s}^{t}} {\min}
			& &  \sum_{i \in \mathcal{N}}
			~ \Big( r_{i}^{t}(\boldsymbol{s}) \Big) ^{2} C_{i} \\
			& \text{subject to}
			& & \text{Constraints \eqref{s_constraint1}, \eqref{s_constraint2} and \eqref{energy_constraint2}}.
		\end{aligned}
	\end{equation*}
	\begin{align*}
		\leftline{\textbf{RS2: Stage 2 of the Restricted Problem}}
	\end{align*}
	\begin{equation*}
		\begin{aligned}
			& \underset{\boldsymbol{e}^{t}} {\min}
			& & \sum_{i \in \mathcal{N}}
			\Big( \alpha_{i}^{t} +\beta_{i} (e_{i}^{t}-s_{i}^{t}) \Big) e_{i}^{t} \\
			& \text{subject to}
			& & \text{Constraints \eqref{energy_constraint1}}.
		\end{aligned}
	\end{equation*}
	
	We use backward induction to solve \textbf{RS1} and \textbf{RS2}. We first solve Problem \textbf{RS2}. Since \textbf{RS2} is a convex quadratic program with equality constraints, we obtain the optimal solution in the closed form as
	\begin{equation}
	e_{i}^{t}= \frac {s_{i}^{t}}{2} - \frac {\alpha_{i}^{t}+ \theta_i \sigma^t} {2  \beta_{i}},~\forall i \in \mathcal{N}, \label{RSsolution}
	\end{equation}
	where $\sigma^{t}$ is the Lagrangian multiplier corresponding to the energy equality constraint \eqref{energy_constraint1}.
	
	Substituting the optimal solution of Problem \textbf{RS2} \eqref{RSsolution} into Problem \textbf{RS1}, we have the restricted problem as a single-level optimization problem:
	\begin{align*}
		\leftline{\textbf{RS: Restricted Problem}}
	\end{align*}
	\begin{equation*}
		\begin{aligned}
			& \underset{ \{ \boldsymbol{s}^{t}, \boldsymbol{e}^{t}, \sigma^t \}} {\min}
			& & \sum_{i \in \mathcal{N}}
			~  \Big( r_{i}^{t} \Big) ^{2} C_{i} \\
			& \text{subject to}
			& & \text{Constraints \eqref{s_constraint1}--\eqref{RSsolution}},
		\end{aligned}
	\end{equation*}
	which is a convex quadratic program, and can be solved by standard convex programming algorithms \cite{convex}.

	Intuitively, moving constraints \eqref{energy_constraint2} from the Stage-2 problem to the Stage-1 problem shrinks the utility company's action set. This can lead to a performance degradation in term of a higher \emph{ELI}, which serves as a upper bound stated in the following proposition.
	\begin{proposition}\label{upperbound}
		The optimal solution of the restricted problem \textbf{RS} provides an upper bound of the optimal \emph{ELI} performance of \textbf{PS1}.
	\end{proposition}
	
	To prove \emph{Proposition 2}, we need to the show that the feasible set of the restricted problem \textbf{RS} is smaller than that of the original two-stage optimization problem. Notice that the constraints \eqref{energy_constraint2} were in \textbf{PS2} of the original two-stage problem formulation, but here we move them to the Stage-1 problem \textbf{RS1} of the restricted two-stage formulation. Compared with \textbf{PS1}, in \textbf{RS1} the utility company's pricing decision in the restricted problem is more conservative, because the utility company has to satisfy the additional data center operation constraints \eqref{energy_constraint2}. Intuitively, the restricted two-stage problem has a smaller feasible set than that of the original two-stage problem. Therefore, the solution that is obtained from the restricted problem provides an upper bound for the original two-stage problem. For detailed proof, see Appendix C.
	
	Note that the estimation of the upper bound is affected by the parameter configurations in constraints \eqref{s_constraint1}, \eqref{s_constraint2} and \eqref{energy_constraint2}.

	\section{Solving the Original Two-stage Problem}
	After presenting \emph{ELI} performance upper and lower bounds from the benchmark problems, we next solve the original two-stage problem through backward induction. We first solve the Stage-2 problem \textbf{PS2}, where data centers minimize the total energy cost. Then, we design a branch-and-bound algorithm for the Stage-1 problem \textbf{PS1} to attain the globally optimal solution.

	\subsection{Solving the Stage-2 Problem}
	In the Stage-2 problem \textbf{PS2}, data centers decide the workload allocation $\lambda_{i}^{t}$ and number of active servers $x_{i}^{t}$ at all locations to minimize the total energy cost in each time slot, given the charging reference $\boldsymbol{s}^{t}$ announced by the utility company ahead of each time slot. 
	
	We have reformulated \textbf{PS2} as en equivalent problem \textbf{PE2} in Section III. As Problem \textbf{PE2} is strictly convex, we can compute the optimal solution $e_i^{t*}$ through the Lagrangian dual method. This leads to the following result.
	\begin{theorem}\label{solve_pe2}
		The unique optimal solution of Problem \textbf{PE2} is
		\begin{equation}\label{opt_pe2-t}
			e_{i}^{t*}(\boldsymbol{s}^{t})= \min
			\left\{
			\max \left\{ \underline{E}_{i}^{t},
			\frac {s_{i}^{t}}{2} - \frac {\alpha_{i}^{t}+ \theta_i \sigma^t} {2 \beta_{i}} \right\}
			,\overline{E}_{i}^{t} \right\},~\forall i \in \mathcal{N}.
		\end{equation}
	\end{theorem}
	where $e_{i}^{t \ast}(\boldsymbol{s}^{t})$ is called the best response of data center $i$ to the billing reference $\boldsymbol{s}^{t}$, and $\sigma^t$ is the Lagrangian multiplier corresponding to the equality constraint \eqref{energy_constraint1}. 
	
	Problem \textbf{PE2} can be solved by the standard subgradient method with a constant stepsize \cite{convex}. For detailed proof, see Appendix D.

	\subsection{Solving the Stage-1 Problem}
	After solving the Stage-2 problem \textbf{PE2}, we obtain the optimal energy consumption of data centers as functions of the given charging references $\boldsymbol{s}^{t}$. We next solve the Stage-1 problem \textbf{PS1}. Under the assumption of complete information, the utility company knows how the data centers will respond to the dynamic prices, and can predict the energy consumptions of data centers given the dynamic prices.\footnote{We assume that the utility company can predict the energy consumptions of data centers through long-term observation, as the utility company is the electricity provider and knows the historical energy consumptions of all the data centers.} Therefore, we can replace Problem \textbf{PE2} with its Karush-Kuhn-Tucker (KKT) conditions and transform the two-stage problem to a single-level optimization problem \cite{kktmethod} by incorporating the KKT conditions of Problem \textbf{PE2} into Problem \textbf{PS1}.
	\begin{theorem}\label{reformulate_lp}
		(Reformulation) The Stage-1 problem \textbf{PS1} can be written in the following equivalent problem with quadratic objectives, linear constraints, and complementarity constraints, denoted as \textbf{PE1}.
	\end{theorem}
	\begin{align*}
		\leftline{\textbf{PE1: Equivalent Problem of the Two-stage Problem}}
	\end{align*}
	\begin{align}
		& \min_{ \{ s_i^t,e_i^t,\sigma^t,\underline{\omega}_i^t,\overline{\omega}_i^t \}, i \in \mathcal{N} }   
		\begin{aligned}[t]
			\sum_{i \in \mathcal{N}} (r_i^t)^{2} C_i    
		\end{aligned} \notag \\
		& \text{subject to} \notag \\
		& \underline{\pi}_i^t \leq \alpha_{i}^{t} + \beta_{i} (e_{i}^{t}-s_i^t) \leq \overline{\pi}_i^t,~\forall i \in \mathcal{N}, \label{pe1-t-1} \\
		& \frac{1}{N} \sum_{i \in \mathcal{N}} \left[ \alpha_{i}^{t} + \beta_{i} (e_{i}^{t}-s_i^t) \right] \leq \pi_{\max}^{t}, \label{pe1-t-2}\\
		& \alpha_{i}^{t}+2 \beta_{i} e_{i}^{t} - \beta_{i} s_{i}^{t} + \theta_i \sigma^{t} - \underline{\omega}_{i}^{t} + \overline{\omega}_{i}^{t} =0,~\forall i\in\mathcal{N}, \label{pe1-t-3} \\
		& \underline{\omega}_{i}^{t} (\underline{E}_{i}^{t} - e_{i}^{t})=0,~\forall i\in\mathcal{N}, \label{pe1-t-4} \\
		& \overline{\omega}_{i}^{t} (e_{i}^{t} - \overline{E}_{i}^{t}) =0,~\forall i\in\mathcal{N}, \label{pe1-t-5} \\
		& \sum_{i \in \mathcal{N}} \theta_i e_{i}^{t} = E^{t}, \label{pe1-t-6} \\
		& \underline{E}_{i}^{t} \leq e_{i}^{t} \leq \overline{E}_{i}^{t},~\forall i\in\mathcal{N}, \label{pe1-t-7} \\
		& \underline{\omega}_{i}^{t} \geq 0,~\overline{\omega}_{i}^{t} \geq 0,~~\forall i\in\mathcal{N}, \label{pe1-t-8} 
	\end{align}
	where \eqref{pe1-t-3}-\eqref{pe1-t-8} are the KKT conditions of Problem \textbf{PE2}, and $\sigma^t$, $\underline{\omega}_i^t$, and $\overline{\omega}_i^t$ are the Lagrange multipliers associated with the equality and box constraints of \textbf{PE2}. Since Problem \textbf{PE2} is strictly convex, the KKT conditions \eqref{pe1-t-3}-\eqref{pe1-t-8} are necessary and sufficient for the optimal solution of Problem \textbf{PE2}.
	
	Problem \textbf{PE1} is a quadratic program with nonconvex constraints, which cannot be solved efficiently by standard convex optimization techniques. However, we find that the nonconvexity only comes from the complementarity slackness conditions \eqref{pe1-t-4} and \eqref{pe1-t-5}. We can linearize the complementarity slackness conditions \eqref{pe1-t-4} and \eqref{pe1-t-5} by introducing binary variables $\underline{z}_{i}^{t} \in \{ 0,1 \}$ and $\overline{z}_{i}^{t} \in \{ 0,1 \}$, and replace \eqref{pe1-t-4} and \eqref{pe1-t-5} by the following constraints:
	\begin{align}
		& e_{i}^{t} - \underline{E}_{i}^{t} \leq \underline{z}_{i}^{t} K ,~\forall i\in\mathcal {N}, \label{linearize-1} \\
		& \underline{\omega}_{i}^{t} \leq (1-\underline{z}_{i}^{t}) K ,~\forall i\in\mathcal {N}, \label{linearize-2} \\
		& \overline{E}_{i}^{t} - e_{i}^{t} \leq \overline{z}_{i}^{t} K ,~\forall i\in\mathcal {N}, \label{linearize-3} \\
		& \overline{\omega}_{i}^{t}  \leq (1-\overline{z}_{i}^{t}) K ,~\forall i\in\mathcal {N}, \label{linearize-4}
	\end{align}
	where $K$ is a sufficiently large constant. We can show that \eqref{pe1-t-4} is equivalent to \eqref{linearize-1} and \eqref{linearize-2}. 
	\begin{itemize}
		\item We first show that if \eqref{pe1-t-4} is satisfied, then (23) and (24) are also satisfied. There are three combinations to make \eqref{pe1-t-4} be satisfied. 1) When $e_{i}^{t} = \underline{E}_{i}^{t}$ and $\underline{\omega}_{i}^{t} > 0$, we have $\underline{z}_{i}^{t} \in [0, 1-\frac{\underline{\omega}_{i}^{t}}{K}]$ from \eqref{linearize-1} and \eqref{linearize-2}. As $\underline{z}_{i}^{t}$ is a binary variable, we obtain that $\underline{z}_{i}^{t} =0$. 2) When $e_{i}^{t} > \underline{E}_{i}^{t}$ and $\underline{\omega}_{i}^{t} = 0$, we obtain that $\underline{z}_{i}^{t} = 1$. 3) When $e_{i}^{t} = \underline{E}_{i}^{t}$ and $\underline{\omega}_{i}^{t} = 0$, we obtain that $\underline{z}_{i}^{t} \in [0,1]$, and thus either $\underline{z}_{i}^{t} = 0$ or $\underline{z}_{i}^{t} = 1$.
		\item We then show that if \eqref{linearize-1} and \eqref{linearize-2} are satisfied, then \eqref{pe1-t-4} is also satisfied. We discuss the following two cases by exhausting the choices of the binary variable $\underline{z}_{i}^{t}$. 1) When $\underline{z}_{i}^{t} = 0$, we have $e_{i}^{t} \leq \underline{E}_{i}^{t}$ from \eqref{linearize-1}. Together with the constraint $e_{i}^{t} \geq \underline{E}_{i}^{t}$ as in \eqref{pe1-t-7}, we obtain $e_{i}^{t} = \underline{E}_{i}^{t}$, and thus \eqref{pe1-t-4} is satisfied. 2) When $\underline{z}_{i}^{t} = 1$, we have $\underline{\omega}_{i}^{t} \leq 0$ from \eqref{linearize-2}. Together with the constraint $\underline{\omega}_{i}^{t} \geq 0$ as in \eqref{pe1-t-8}, we have $\underline{\omega}_{i}^{t} = 0$, and thus \eqref{pe1-t-4} is also satisfied. 
	\end{itemize}
		Following a similar reasoning, we can show that (\ref{linearize-3}) and (\ref{linearize-4}) can replace (\ref{pe1-t-5}).
	
	To solve \textbf{PE1}, we design a branch-and-bound algorithm \cite{bilevel} to attain the optimal solution. We first relax the binary variables $\{ 0,1 \}$ to continuous variables within the range $[0,1]$, and define the following relaxed quadratic problem \textbf{PR1}.
	\begin{align*}
		\leftline{\textbf{PR1: Relaxed Problemof PE1}}
	\end{align*}
	\begin{equation*}
		\begin{aligned}
			&\min  
			& & \sum_{i \in \mathcal{N}} ~\Big( r_i^t \Big)^{2} C_i\\
			& \text{subject to}
			& & \text{Constraints}~ \eqref{pe1-t-1}-\eqref{pe1-t-3},~\eqref{pe1-t-6}-\eqref{linearize-4}, \\
			&&& 0 \leq \underline{z}_{i}^{t} \leq 1,~\forall i\in\mathcal {N},\\
			&&& 0 \leq \overline{z}_{i}^{t} \leq 1,~\forall i\in\mathcal {N},\\
			& \text{Variables:} 
			& & \{ s_i^t,e_i^t,\sigma^t,\underline{\omega}_i^t,\overline{\omega}_i^t, \underline{z}_{i}^{t}, \overline{z}_{i}^{t} \}, i \in \mathcal{N}.
		\end{aligned}
	\end{equation*}

\begin{figure}[t]
		\centering
		\includegraphics[width=5cm]{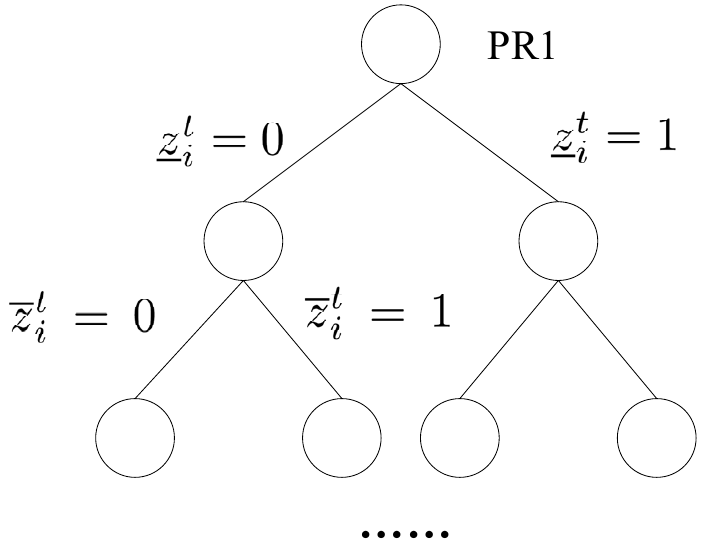}
		\caption{\label{fig_bbtree}Branch and bound tree.}
	\end{figure}

	Initially, the algorithm takes the optimum of the integrated problem \textbf{PI} as the lower bound $\underline{F}$, and the optimum of the restricted problem \textbf{RS} as the upper bound $\overline{F}$. Then the algorithm start to solve the relaxed problem \textbf{PR1} and builds the branch and bound tree by splitting the binary variables to enforce the binary variable constraints. Specifically, the algorithm adds the following constraints, $\underline{z}_{i}^{t}=0$ or $ \underline{z}_{i}^{t}=1$, to the relaxed problem \textbf{PR1}, and derives two new convex quadratic problems (\textit{e.g.}, two first-level children nodes in the branch-and-bound tree shown in Fig. \ref{fig_bbtree}). The algorithm continues to expand the tree by adding other constraints $\overline{z}_{i}^{t}=0$ or $ \overline{z}_{i}^{t}=1$ until all the binary variables constraints are completely enforced. Meanwhile, the algorithm updates the lower bound $\underline{F}$ after solving each relaxed problem in the children node, and updates the upper bound $\overline{F}$, when a feasible solution with lower optimum is found. The branch-and-bound algorithm terminates at a globally optimal solutions when the lower bound meets the upper bound or all the nodes in the branch and bound tree have been evaluated \cite{bilevel}. In the worst-case, the branch-and-bound algorithm will traverse $2^{2N}$ nodes.
	
	\subsection{Heuristic Algorithm}
	The branch-and-bound algorithm in general has a very high worst-case computational complexity, and hence may not be suitable for solving a large-scale load balancing problem. Therefore, we propose a heuristic algorithm to solve the two-stage problem \textbf{PS1} and \textbf{PS2} for suboptimal solutions. Our heuristic algorithm is designed based on the descent approach, and iteratively reduces the value of \emph{ELI} in Problem \textbf{PS1}.
	
	We view the solution of Problem \textbf{PE2} as a function of the variables of Problem \textbf{PS1}. Observing the best response (\ref{opt_pe2-t}) in Problem \textbf{PE2}, we find the following monotonic relation between the charging reference $s_i^t$, the optimal energy consumption $e_i^{t \ast}$, and the unit price $\pi_i^t = \alpha_i^t + \beta_i (e_i^{t \ast} - s_i^t)$. Specifically, increasing $s_i^t$ leads to increase in $e_i^{t \ast}$ and $\sigma^t$, and decrease in $e_j^{t \ast},~\forall j \in \mathcal{N} \backslash i$, and all the unit prices $\pi_i^t$ also decrease. On the contrary, decreasing $s_i^t$ causes decrease in $e_i^{t \ast}$ and $\sigma^t$, and increase in $e_j^{t \ast},~\forall j \in \mathcal{N} \backslash i$ and all the unit prices $\pi_i^t$. Note that minimizing the \emph{ELI} performance (\ref{def_eli}) yields even load ratio $r_i^t$ across all locations. Thus, we design a descent algorithm to redistribute the total energy consumptions, by decreasing $s_i^t$ in high energy-consumption locations, and increasing $s_i^t$ in low energy-consumption locations. The detailed algorithm is described in Algorithm 1. The utility company and data center iteratively compute the prices and energy consumption. In each iteration, the utility company provides a set of prices, and data centers respond to the prices and report the corresponding schedule of energy consumption (and do not reveal private information such as parameters and constraints). Algorithm 1 reduces \emph{ELI} and its convergence to a feasible and possibly sub-optimal solution is guaranteed since the \emph{ELI} performance is lower bounded by Problem \textbf{PI}. For detailed proof, see Appendix E.
	
	\begin{algorithm}[h]
		\caption{Descent algorithm to solve the two-stage problem}
		\label{alg1}
		\begin{algorithmic}[1]
			\State \textbf{Initialization}: In each time slot $t \in \{1,...,T \}$, set the iteration count $k=1$, convergence tolerance $\epsilon>0$, and step-size $\eta(k)$. Initialize the starting point $\boldsymbol{s}^t(k) \triangleq \{ s_i^t(k),i\in \mathcal{N}\}$ by solving the restricted problem \textbf{RS}, and compute the average load ratio $r_{avg}^t(k) = \frac{\sum_{i \in \mathcal{N}} r_i^t(k)}{N} $. 
			\Repeat
			\State \textbf{Step1:} Compute the descent direction $\boldsymbol{g}^t (k)$ for $\boldsymbol{s}^t (k)$: if $r_i^t(k) > r_{avg}^t (k)$, then set $g_{i}^t (k)= - \frac{\theta_{i}}{\beta{i}}$, $i \in \mathcal{N}$; otherwise, set $g_{j}^t (k)= \frac{\theta_{j}}{\beta{j}}$, $j \in \mathcal{N} \backslash i$.
			
			\State \textbf{Step2:} Perform the search by using the iterations
			\State $\boldsymbol{s}^t (k+1)=\boldsymbol{s}^t (k) + \eta(k) \boldsymbol{g}^t(k)$;
			
			\State \textbf{Step3:} Given $\boldsymbol{s}^t(k+1)$, solve the optimal energy consumption $\boldsymbol{e}_{i}^{t}(k+1)$ according to (\ref{opt_pe2-t}).
			
			\State \textbf{Step4:} Check the feasibility based on (\ref{s_constraint1}) and (\ref{s_constraint2}). If yes, update $r_{avg}^t(k+1) = \frac{\sum_{i \in \mathcal{N}} r_i^t(k+1)}{N} $. 
			If not, 
			\begin{flalign*} 
				& e_{i}^{t}(k+1)=e_{i}^{t}(k),~s_{i}^{t}(k+1)=s_{i}^{t}(k),\\
				& r_{avg}^t(k+1)=r_{avg}^t(k),~\eta(k+1) = \frac{1}{2} \eta(k).
			\end{flalign*}
			
			\State $k \gets k+1$;
			
			\Until the convergence criteria $\| ELI(k) - ELI(k-1) \| \leq \epsilon$ is satisfied;
			\State Return the sub-optimal solutions $\boldsymbol{\hat{s}}^{t}$, $\boldsymbol{\hat{e}}^{t}$.
			\State \textbf{end}
		\end{algorithmic}
	\end{algorithm}

	\section{Impact of Background Load Prediction Error}
	In Section V, we solved the two-stage problem based on the assumption that the utility company can forecast the background power load $B_i^t$ accurately. In practice, the prediction may have errors and the actual background load may deviate from the predicted values. We define the prediction errors for the background load in location $i$ and time slot $t$ as $\delta_{i}^{t}$. Then, we can represent the actual background load $\hat{B}_{i}^{t}$ as the summation of predicted value and the prediction error:
	\begin{align*}
	\hat{B}_{i}^{t} = B_{i}^{t} + \delta_{i}^{t}. 
	\end{align*}
	
	Next we use the robust optimization approach \cite{worst-case} to analyze the impact of prediction errors. We assume that the prediction errors are bounded in known uncertainty sets as follows:
	\begin{align}
	&  \Delta_{i,\min}^{t} \leq \delta_{i}^{t} \leq \Delta_{i, \max}^{t},~\forall i \in \mathcal{N}, \label{loaderror}
	\end{align}
	where $\Delta_{i,\min}^{t}$ and $\Delta_{i, \max}^{t}$ denote the lower bound and upper bound of the background load prediction error in location $i$ and time slot $t$, respectively.
	
	We let $\boldsymbol{\delta}^{t} = \{ \delta_{i}^{t},~ i \in \mathcal{N} \}$ denote the prediction-error vector. Our aim is to maximize the worst-case performance of power load balancing. We formulate the worst-case performance optimization problem as:
	\begin{align*}
	& \leftline{\textbf{WCP: Worst-case Performance Optimization Problem}}
	\end{align*}
	\begin{equation*}
	\begin{aligned}
	& \underset{\boldsymbol{s}^{t}} {\min} ~\max_{\boldsymbol{\delta}^{t}}
	& & \sum_{i \in \mathcal{N}}~~   \frac{ \Big( e_{i}^{t}(\boldsymbol{s}^{t})+B_{i}^{t} + \delta_{i}^{t} \Big)^{2} }{C_i} \\
	& \text{subject to}
	& & \text{Constraints \eqref{s_constraint1}, \eqref{s_constraint2}, \eqref{loaderror}},
	\end{aligned}
	\end{equation*}
	which is a min-max optimization problem. 
		
	To solve Problem \textbf{WCP}, we first solve the inner \emph{ELI} maximization problem of \textbf{WCP} (namely \textbf{IWCP}):
	\begin{equation*}
	\begin{aligned}
	& \max_{\boldsymbol{\delta}^{t}}
	& & \sum_{i \in \mathcal{N}}~~   \frac{ \Big( e_{i}^{t}(\boldsymbol{s}^{t})+B_{i}^{t} + \delta_{i}^{t} \Big)^{2} }{C_i} \\
	& \text{subject to}
	& & \text{Constraints \eqref{loaderror}},
	\end{aligned}
	\end{equation*}	
	which corresponds to the worst-case \emph{ELI} performance. We can show that the objective function of Problem \textbf{IWCP} is convex in the prediction errors $\boldsymbol{\delta}^{t}$. Hence, the optimal solution of the \textbf{IWCP} problem must reach the boundary of the uncertainty set in \eqref{loaderror}. Moreover, as the total actual energy consumption $e_{i}^{t}(\boldsymbol{s}^{t})+B_{i}^{t} + \delta_{i}^{t}$ is always non-negative, hence the objective function of Problem \textbf{IWCP} is a monotonically increasing function in $\boldsymbol{\delta}^{t}$. Thus we have the following result:
	\begin{proposition}
		The optimal solution of Problem \textbf{IWCP}, \emph{i.e.}, the worst-case prediction error, lies at the upper bounds of the uncertainty set, \emph{i.e.} $\delta_i^{t,\ast} = \Delta_{i,\max}^{t},~\forall i \in \mathcal{N} $.
	\end{proposition}

    Hence, we substitute the worst-case prediction error $\boldsymbol{\delta}^{t,\ast} = \{\delta_i^{t,\ast},~\forall i \in \mathcal{N}\}$ into Problem \textbf{WCP}, and obtain the following worst-case optimization problem:
    \begin{equation*}
    \begin{aligned}
    & \underset{\boldsymbol{s}^{t}} {\min} 
    & & \sum_{i \in \mathcal{N}}~~   \frac{ \Big( e_{i}^{t}(\boldsymbol{s}^{t})+B_{i}^{t} + \delta_i^{t,\ast} \Big)^{2} }{C_i} \\
    & \text{subject to}
    & & \text{Constraints \eqref{s_constraint1} and \eqref{s_constraint2}},
    \end{aligned}
    \end{equation*}
    which solves the optimal billing references $\boldsymbol{s}^{t}$ to optimize the worst-case performance of \emph{ELI}. Note that the above problem shares the same structure as Problem \textbf{PS1}, and thus can be solved by the same methodology presented in Section V.

	\section{Simulation Results}
	In this section, we evaluate our proposed algorithms based on realistic system parameters, and compare the corresponding electric load index and energy cost between the solutions with that of the benchmark problems. 
	
	We consider four data centers that are geographically located in four different regions in the United States: New York, Maine, Rhode Island, and Boston. In each location, there is one data center powered by a power station. The numbers of servers in the four locations are 80000, 60000, 60000, and 80000, respectively. The service rates are 4, 3, 4 and 3 requests per server, and each server consumes 200watts electricity in the peak mode and 100watts when it is idle. We set power usage effectiveness as 1.5, 1.2, 1.2 and 1.5 for four data centers, respectively. We took hourly locational marginal prices and demands of the four locations on 4th March 2013 as the base prices and background power load, according to \cite{data1,data2}. The dynamic computing requests are simulated based on the normalized workload trace of Google data centers on 20th December 2013 \cite{data3,data4}.
	
	\subsection{Performance of the proposed algorithms}
\begin{figure}[tbhp]
		\vspace{-4mm}
		\centering
		\includegraphics[width=7.0cm]{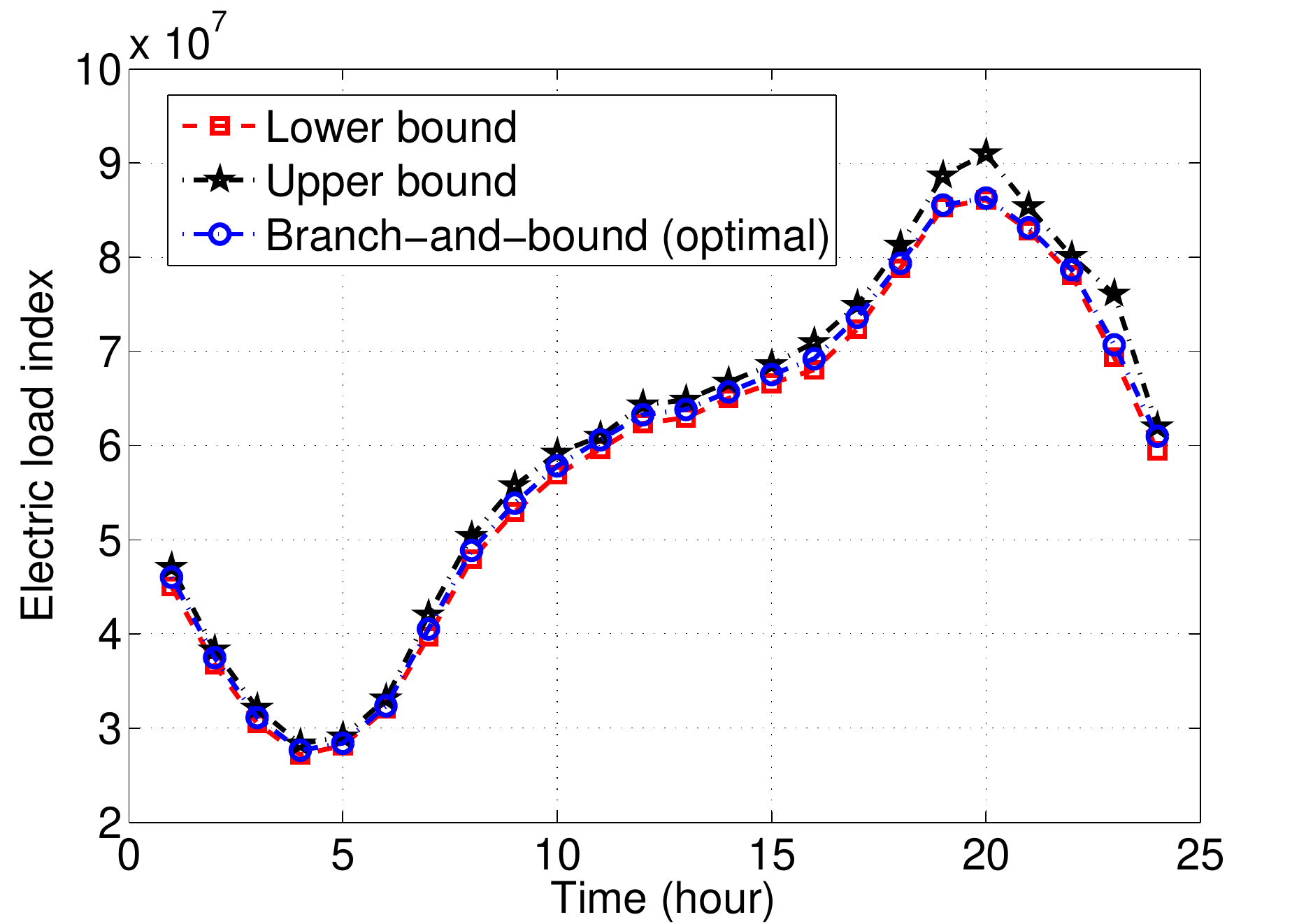}
		\caption{\label{fig_bounds}Upper and lower bounds for ELI.}
		\vspace{-2mm}
	\end{figure}

	We first evaluate the optimal solutions of the two-stage problem and benchmark problems. The upper bound and lower bound for \emph{ELI} over 24 hours are shown in Fig. \ref{fig_bounds}. The optimal solution to the integrated problem provides a lower bound for \emph{ELI}, and the optimal solution to the restricted problem provides an upper bound for \emph{ELI}. Input the upper bound and lower bound into the branch-and-bound algorithm, we can solve the optimal solution to the two-stage problem. The corresponding optimal \emph{ELI} lies in between the upper bound and lower bound, and is very close to the lower bound. Specifically, the optimal \emph{ELI} is on average 1.5\% higher than the lower bound across 24 hours.
	
\begin{figure}[tbhp]
		\vspace{-4mm}
		\centering
		\includegraphics[width=7.0cm]{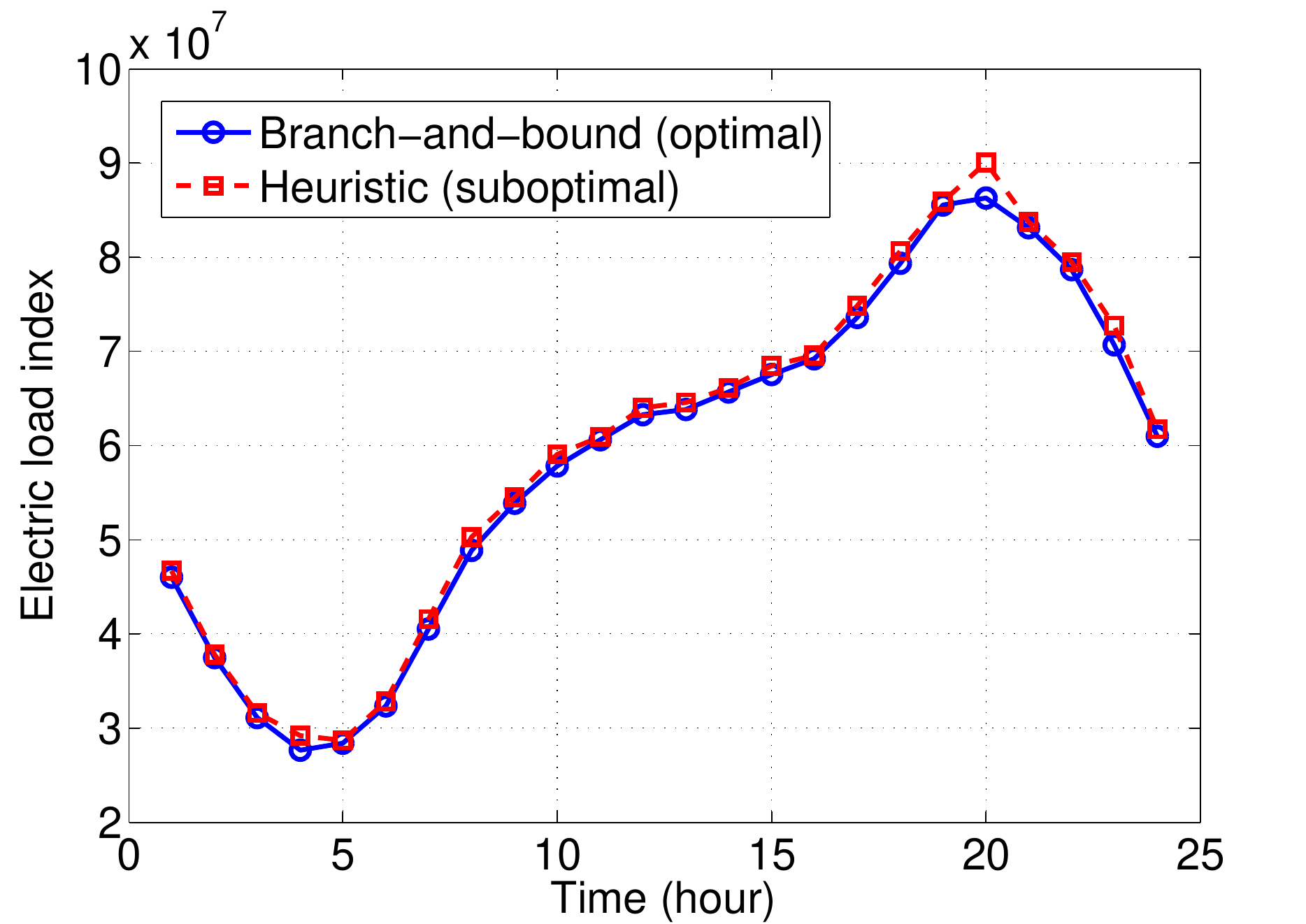}
		\caption{\label{fig_optimal}Optimal and suboptimal ELI.}
		\vspace{-4mm}
	\end{figure}

	In Fig. \ref{fig_optimal}, the solid blue curve represents the optimal \emph{ELI} performance of the branch-and-bound algorithm. The dash red curve represents the sub-optimal \emph{ELI} performance obtained by the heuristic algorithm, which is close to the solid blue curve. This suggests that the heuristic algorithm achieves a performance close to the optimal result. 

	\subsection{Effectiveness of optimized dynamic pricing}
	\begin{figure}[tbhp]
		\vspace{-2mm}
		\centering
		\includegraphics[width=9cm]{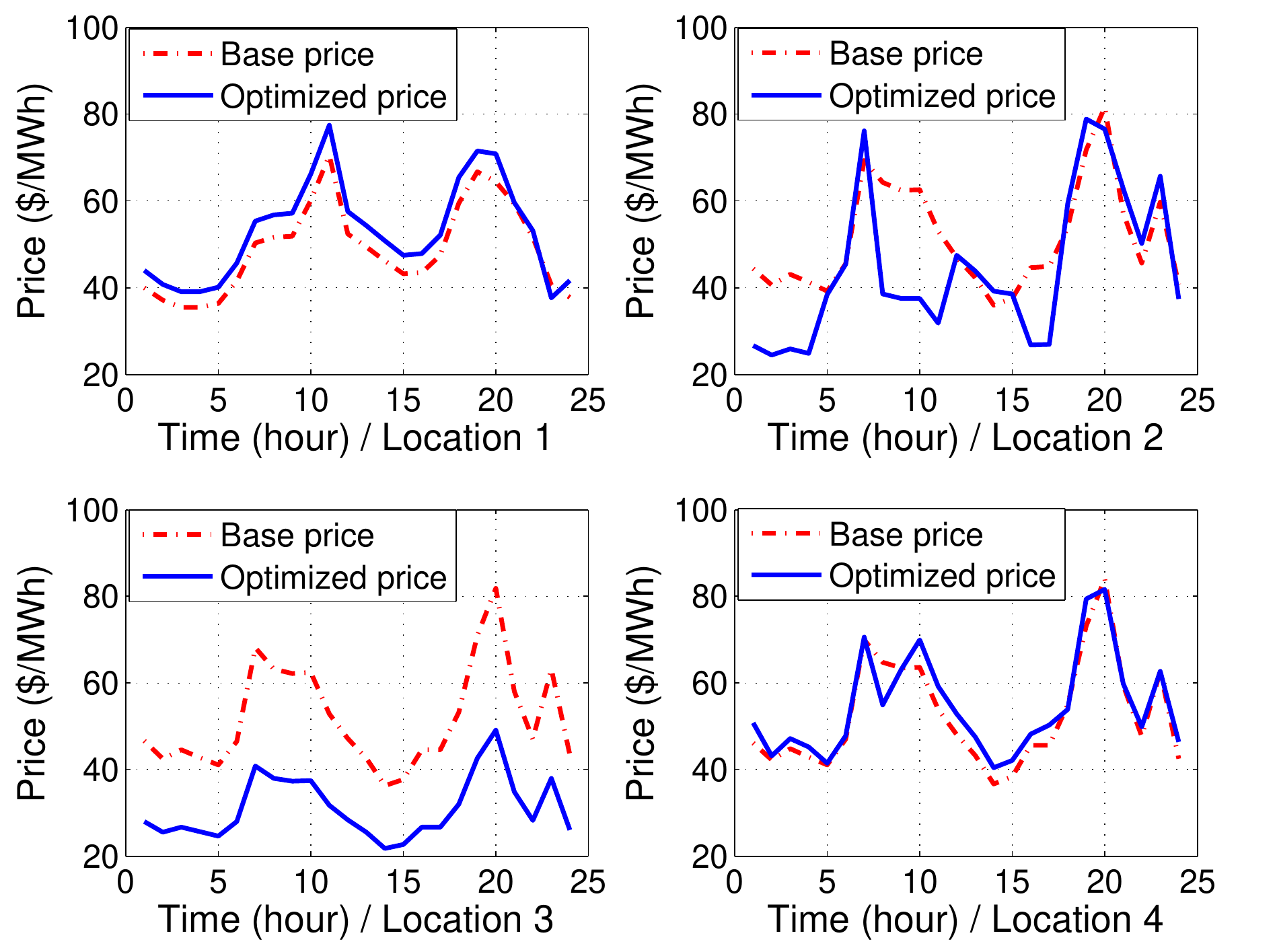}
		\caption{\label{fig_prices}Base pricing and optimized dynamic pricing.}
		\vspace{-2mm}
	\end{figure}
	We demonstrate the effectiveness of our proposed dynamic pricing by comparing to the base pricing benchmark. Fig. \ref{fig_prices} shows the base prices (dash red curves) and the optimized dynamic prices (solid blue curves) for four data centers, respectively. We can see that the optimized prices may significantly deviate from the base prices. Take hour 10 for example, the optimized prices in location 1 and 4 are higher than the base prices, and in location 2 and 3 are lower than the base prices. This implies that the loads in location 1 and 4 are heavier than those in location 2 and 3. The utility company optimizes the prices for the data centers to re-distribute their energy consumption for load balancing.

\begin{figure}[tbhp]
		\vspace{-2mm}
		\centering
		\includegraphics[width=6.8cm]{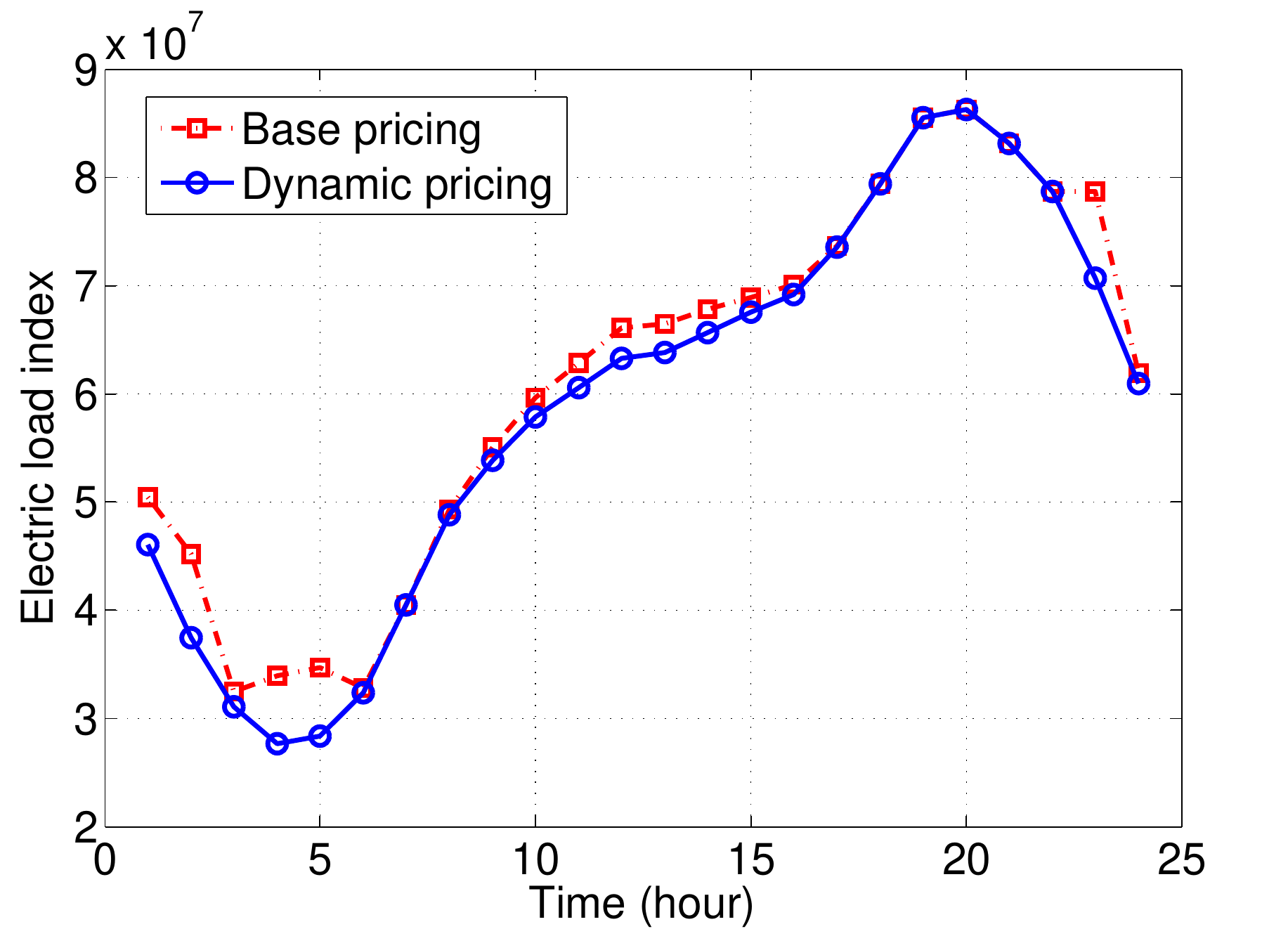}
		\caption{\label{fig_compareeli}Comparison of ELI.}
		\vspace{-4mm}
	\end{figure}

	In Fig. \ref{fig_compareeli}, the dash red curve represents the \emph{ELI} of the base price benchmark, where the data centers are charged based on the fixed base prices. The solid blue curve represents the \emph{ELI} with dynamic pricing, which shows that our proposed dynamic pricing scheme reduces \emph{ELI} by an average of 4$\%$ across 24 hours comparing with the base pricing benchmark.
	
\begin{figure}[tbhp]
		\vspace{-2mm}
		\centering
		\includegraphics[width=7.0cm]{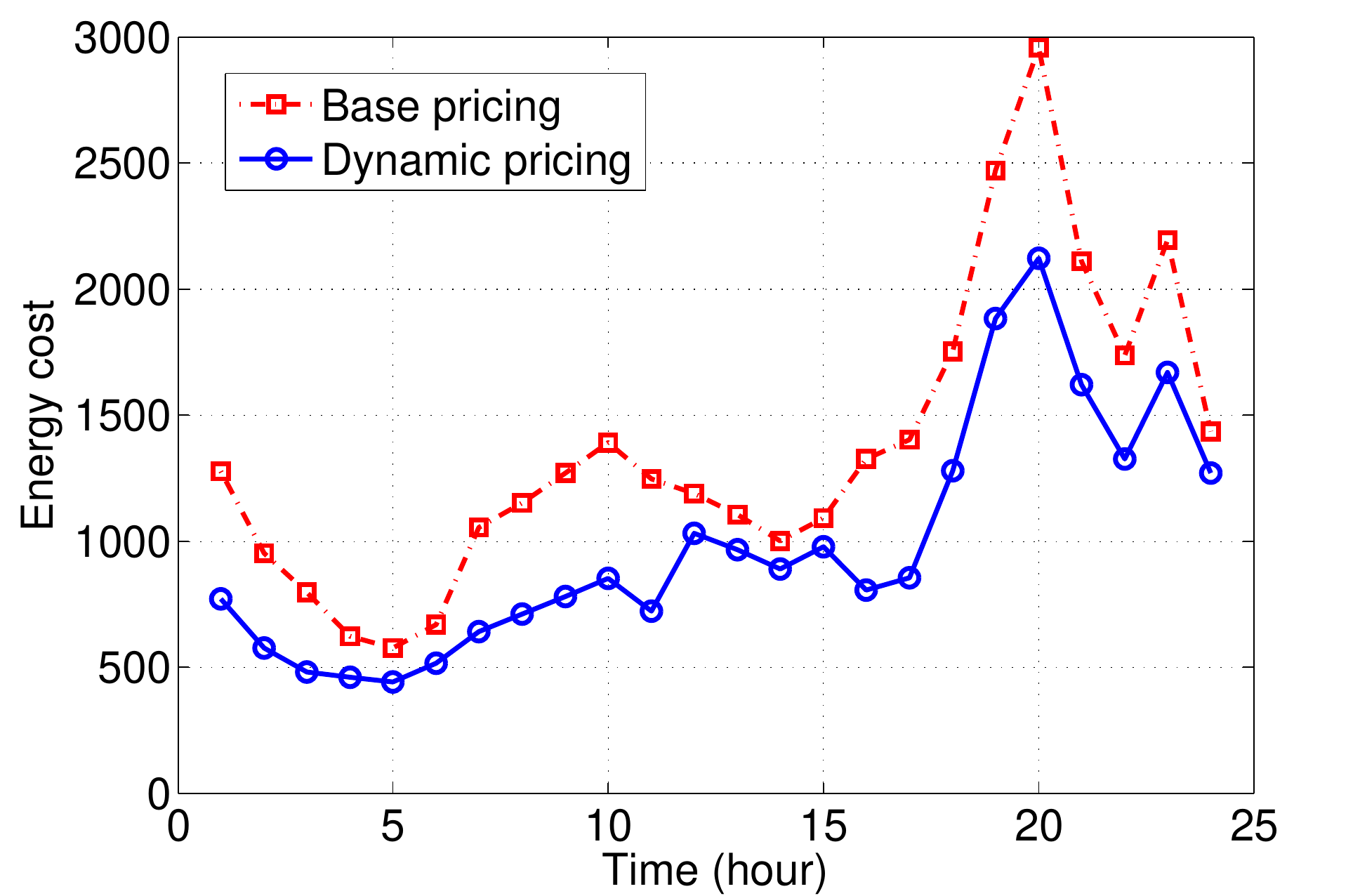}
		\caption{\label{fig_comparecost}Comparison of energy cost.}
		\vspace{-4mm}
	\end{figure}

	We also evaluate data centers' total energy cost over 24 hours, shown in Fig. \ref{fig_comparecost}. The energy cost with dynamic pricing is less than the base pricing benchmark. Specifically, the data centers reduce the total energy cost by an average of 28$\%$ across 24 hours, by taking advantage of dynamic prices and reallocating the workload. Fig. \ref{fig_compareeli} and Fig. \ref{fig_comparecost} show that the dynamic interactions between smart grid and data centers bring benefits to both sides and achieve a win-win situation.

\begin{figure}[tbhp]
		\vspace{-2mm}
		\centering
		\includegraphics[width=7cm]{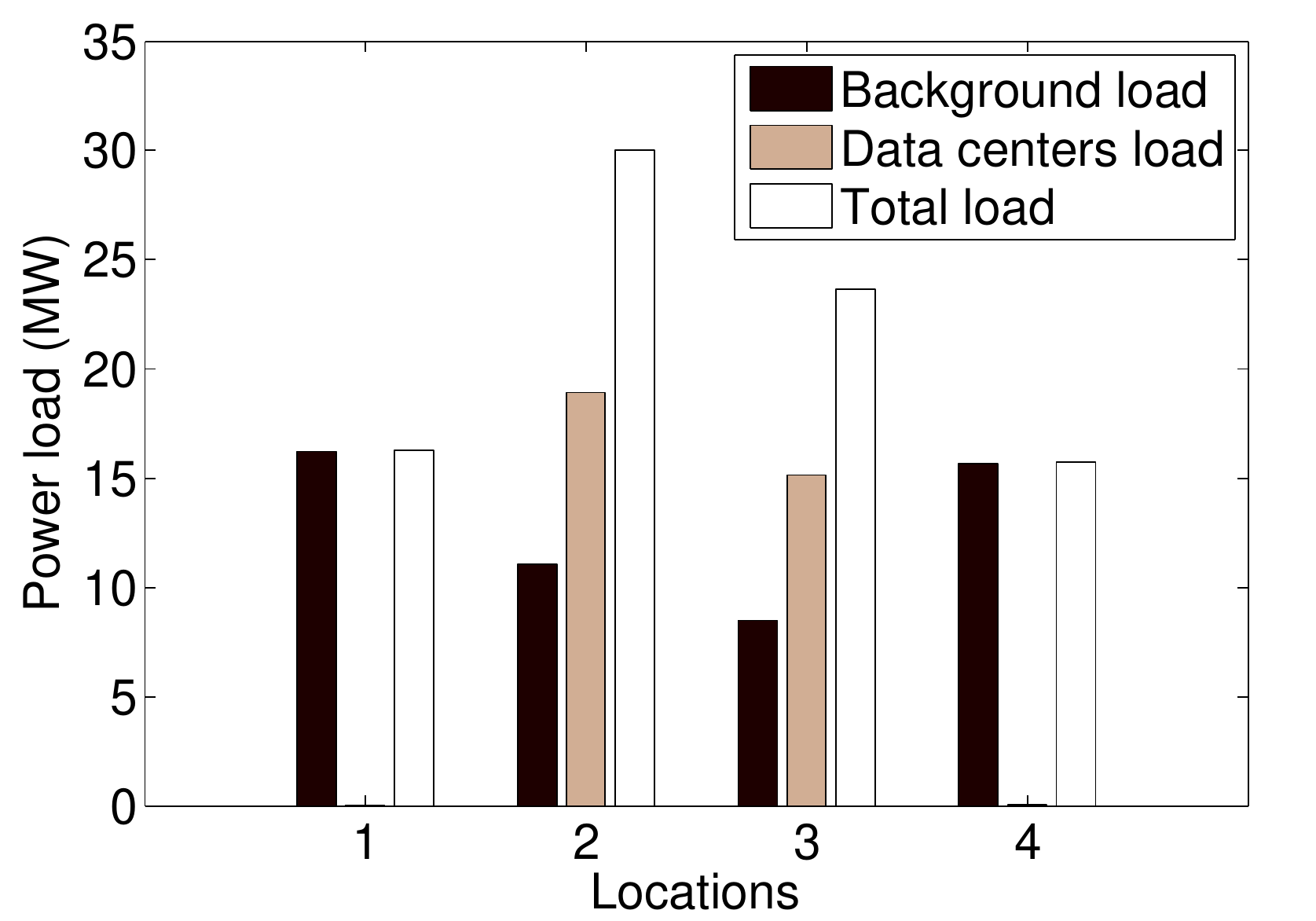}
		\caption{\label{fig_loadbenchmark}Power load (base pricing).}
	\end{figure}

\begin{figure}[tbhp]
		\vspace{-2mm}
		\centering
		\includegraphics[width=7cm]{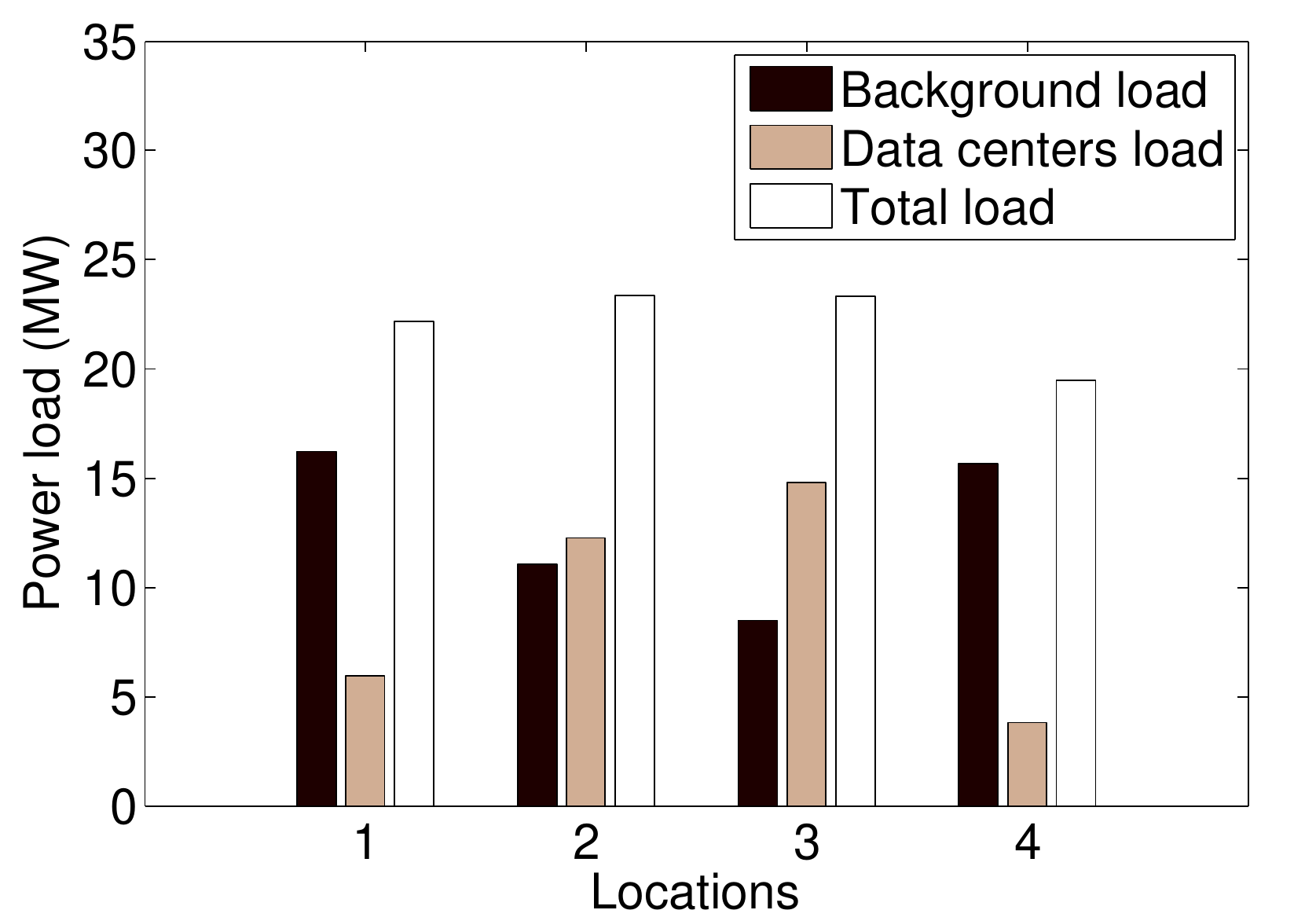}
		\caption{\label{fig_loadoptimal}Power load (dynamic pricing).}
		\vspace{-4mm}
	\end{figure}

	We examine the power load distribution within one particular hour (\emph{e.g.}, hour 24), and plot the background power load, power load of data centers, and the total load across four locations. Fig. \ref{fig_loadbenchmark} shows the load distribution with base pricing. The data centers' load (the white bar) is not balanced, since data centers assign workload to the location with the lowest base price as much as possible to minimize the energy cost. The consequence is that the power load is extremely high in the lowest-price location 2, bringing a risk of overloading. Fig. \ref{fig_loadoptimal} shows the load distribution in the two-stage model with dynamic pricing. We can see the utility company tries to drive the load more evenly across different locations. Therefore, our proposed scheme can effectively improve the reliability of smart grid through re-balancing power load across different locations. 

	\subsection{Impact of prediction errors}
\begin{figure}[tbhp]
		\centering
		\includegraphics[width=7cm]{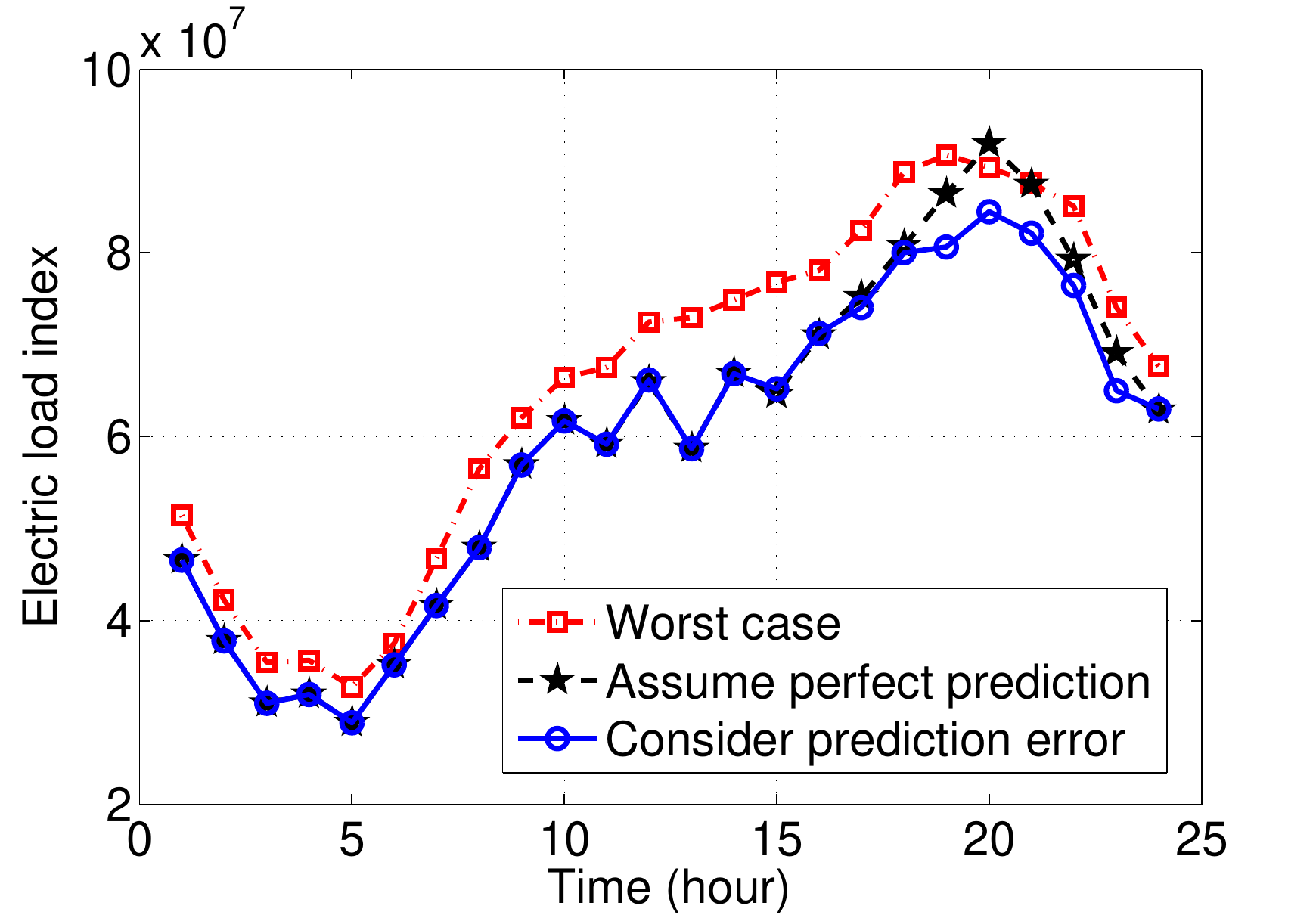}
		\caption{\label{fig_error}ELI performance with prediction error.}
		\vspace{-4mm}
	\end{figure}

	We conduct a case study to show the impact of prediction errors on the \emph{ELI} performance. We set the bounds ($\Delta_{i,\min}^{t}$ and $\Delta_{i,\max}^{t}$) of the prediction errors as $\pm 10 \%$ of the predicted values $B_i^t$ in location $i$ and time slot $t$. Solving problem \textbf{WCP} in Section VI, we obtain the optimized worst-case \emph{ELI} performance as dash red curve in Fig. \ref{fig_error}. We also randomly generate a realization of prediction errors, and compare the \emph{ELI} performance under the scenario with and without considering the prediction errors. If prediction errors are considered when optimizing the Stage-1 problem, the realized \emph{ELI} performance (solid blue curve) can be guaranteed to be better than the worst-case \emph{ELI}. However, if the prediction is assumed to be accurate with zero error (while in reality it is not), then the \emph{ELI} performance (dash black curve) can be even worse than the worst-case benchmark (\emph{e.g.} in the 20th time slot). Therefore, the results demonstrate the effectiveness of our proposed worst-case performance optimization problem, which provides a performance guarantee for \emph{ELI} under prediction errors.

	\subsection{Impact of computing workload}
\begin{figure}[tbhp]
		\vspace{-4mm}
		\centering
		\includegraphics[width=7cm]{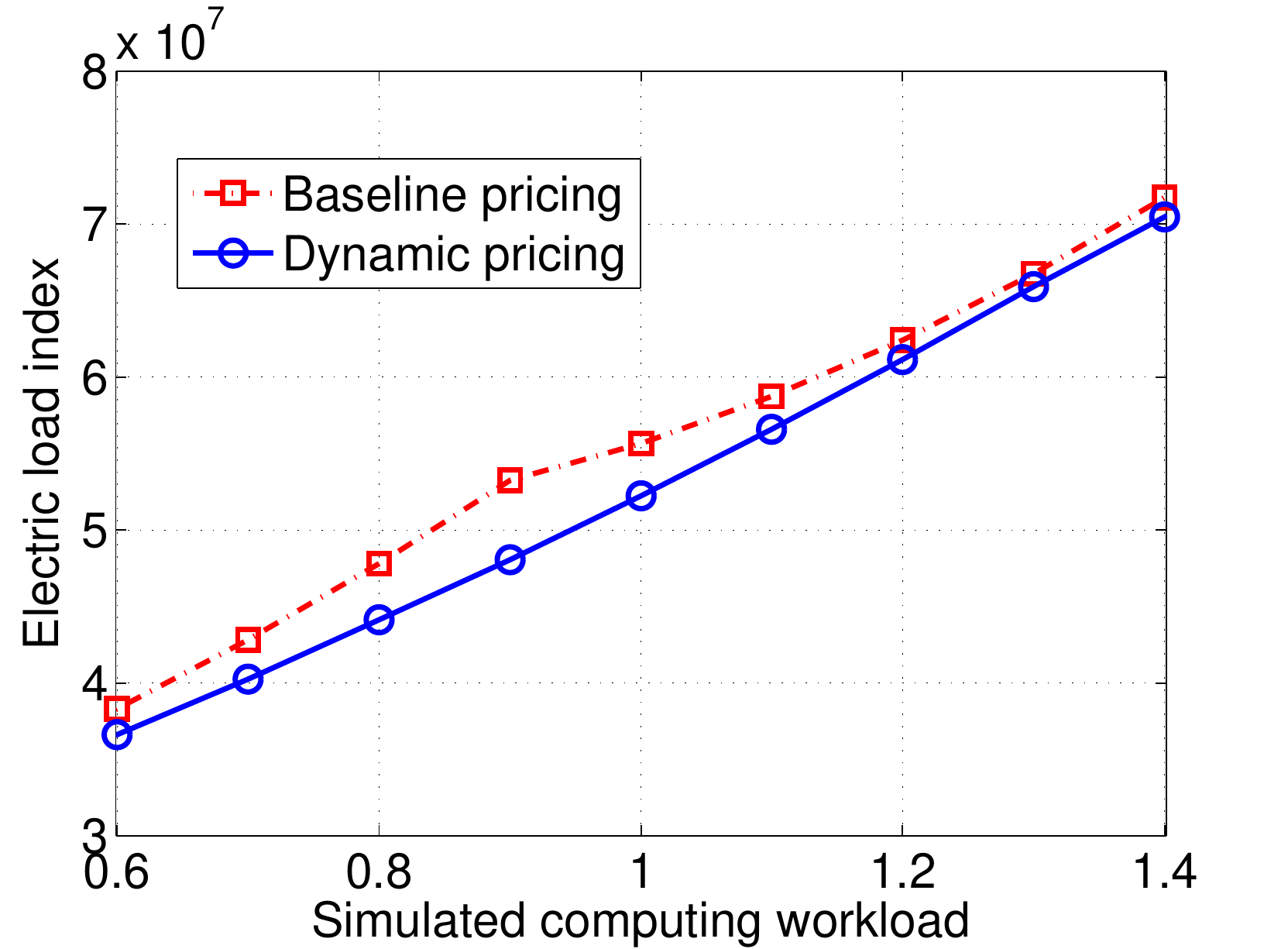}
		\caption{\label{fig_impact_workload} Impact of computing workload.}
		\vspace{-4mm}
	\end{figure}

	We study the impact of computing workload on the performance of dynamic pricing. We consider one time slot (hour 1), and keep all the simulation parameters unchanged except the computing workload. Fig. \ref{fig_impact_workload} depicts the \emph{ELI} performance comparison between the baseline pricing benchmark (fixed price without incentives) and the two-stage optimal dynamic pricing (with incentives), under different computing workload ranging from $0.6$ to $1.4$ of its original value. Fig. \ref{fig_impact_workload} shows that \emph{ELI} increases as the workload increases for both dynamic pricing and baseline pricing, and the two-stage optimal dynamic pricing always achieves a lower \emph{ELI} than the baseline pricing regardless of the computing workload. Moreover, the load balancing improvement (measured by the percentage of \emph{ELI} reduction) is relatively larger under intermediate computing workload than that under light or heavy computing workload. Specifically, dynamic pricing achieves $9.7\%$ in the \emph{ELI} reduction when the workload is $0.9$ of the original value. When the workload is $0.6$ and $1.4$ of the original value, the percentage of \emph{ELI} reduction decreases to $4.4\%$ and $1.8\%$, respectively. The reason is as follows. When the computing workload is light, the corresponding total energy consumption is low, and is not likely to cause power overloading. Therefore, the benefit due to the dynamic pricing is relatively small. When the computing workload is heavy, the cloud provider tends to fully utilize all the data centers to provision the quality of service to all the computing requests. There is little flexibility in terms of shifting workload across different data centers, and thus the benefit of dynamic pricing becomes small as well.

	\section{Conclusions}
	In this paper, we studied the dynamic interactions between smart grid and data centers as a two-stage price optimization problem. To solve the two-stage optimization problem, we reformulated it as a mixed integer quadratic programming problem, and proposed a branch-and-bound algorithm to attain the globally optimal solution, and a low complexity heuristic descent algorithm to yield a close-to-optimal solution. The simulation results showed a win-win solution for both the utility company and data centers.
	
	For future work, we would like to study the interaction between the utility company and data centers with high penetration of renewable energy and under incomplete information. Some cloud provides installed renewable energy facilities to power data centers. How to manage the renewable-powered data centers and what is the impact on the power system are worth of study. The utility company may not able to acquire private information of data-center operation, so how to incentivize data centers with asymmetric information is an interesting and practical problem for future study.

\end{document}